\documentclass[onecolumn]{rsauthor}

\jname{Phil. Trans. R. Soc. A}

\artnum{2011.0549}

\def\gsim{\mathrel{\hbox{\rlap{\hbox{\lower4pt\hbox{$\sim$}}}\hbox{$>$}}}}
\def\lsim{\mathrel{\hbox{\rlap{\hbox{\lower4pt\hbox{$\sim$}}}\hbox{$<$}}}}

\begin{document}

\title{Random time series in Astronomy}

\author{Simon Vaughan$^{1,*}$\thanks{*simon.vaughan@leicester.ac.uk}}

\address{$^1$X-ray \& Observational Astronomy Group, Department of Physics and Astronomy, University of Leicester, Leicester LE1 7RH, UK}

\abstract{Progress in astronomy comes from interpreting the signals encoded in the light received from distant objects: the distribution of light over the sky (images), over photon wavelength (spectrum), over polarization angle, and over time (usually called light curves by astronomers). In the time domain we see transient events such as supernovae, gamma-ray bursts, and other powerful explosions; we see periodic phenomena such as the orbits of planets around nearby stars, radio pulsars, and pulsations of stars in nearby galaxies; and persistent aperiodic variations (`noise') from powerful systems like accreting black holes. I review just a few of the recent and future challenges in the burgeoning area of Time Domain Astrophysics, with particular attention to persistently variable sources, the recovery of reliable noise power spectra from sparsely sampled time series, higher-order properties of accreting black holes, and time delays and correlations in multivariate time series.}


\date{Received 9 March 2012; Revised 04 May 2012; Accepted 08 May 2012}

\maketitle


\section{Introduction}
\label{sect:intro}

Pick up any good book on Time Series Analysis and you will probably find the topics covered include: trends and seasonal components, autoregressive and moving average processes (ARMA), forecasting, Kalman filters, state-space models, spectral methods, and so on. Many of these will sound unfamiliar to astronomers, even those who regularly examine time series in the form of light curves of variable astronomical objects. From the astronomer's perspective time series analysis, as a subject itself, appears to be a branch of statistics but with its own specialist jargon and approach, developed in no small part from applications in signal processing engineering and econometrics. Astronomers generally do not have much formal training in statistics (if any), tend to have their own jargon, and read papers and textbooks only by other astronomers. The exchange of ideas between time series experts and  astronomers with data to analyse is not as effective as it could be. The result is, at best, inefficiency; astronomers spend time struggling with the details of unfamiliar methods or re-inventing techniques that have already been discussed extensively elsewhere. Even between different sub-branches of astronomy (e.g. between X-ray astronomy, optical and radio astronomy; or stellar and extragalactic astronomy) there may be differences of convention and practice (in data collection, curation, analysis and interpretation) that hamper effective communication and transfer of ideas.

As well as the `sociological' barriers that divide astronomers from time series analysts elsewhere, there are more practical reasons that may contribute to their separation. Astronomers often have to deal with limitations that may not trouble data collection in other fields. The most obvious problem is breaks in observations due to observing conditions related to e.g. visibility of the target on the sky, bad weather, or telescope scheduling constraints, resulting in discontinuous and often uneven time sampling. There may be a variable background, measurement errors or instrumental effects (such as `deadtime' or readout cycles) that are beyond experimental control, but must be understood through careful instrumental calibration. Another difference between much of conventional time series analysis and astronomy is their objectives. Much of the time series literature is concerned with modelling time series for the purpose of forecasting (e.g. in meteorology, econometrics and finance, engineering applications). Astronomers are (usually) more interested in developing and testing physical models, to better understand the physical principles at work in a variable system, rather than in predicting future values of some time varying function for their own worth.

The main purpose of this article is, in the spirit of this meeting, to facilitate discussion and the exchange of ideas between those who work on astronomical problems involving variable objects, and those who work on methods for analysing time series data. If, as a result, a few astronomers go on to consider alternative methods for analysing their data, or a few time series specialists from the statistics or signal processing communities get involved with astronomical data, I will consider this to have been a success.


\section{Why time series analysis matters to astronomers}

Astronomy is an observational science. Unlike experimental sciences we are not able to interact with the objects we study, alter experimental conditions or repeat experiments under the same conditions to obtain more data. The information we can gather from astronomical objects (planets, stars, galaxies, etc.) is usually limited to the light we receive from them. (I will not deal here with other types of radiation, e.g. neutrino, cosmic ray and gravitational wave astronomy. The latter topic is discussed elsewhere in this volume.) In principle we are able to examine  how this light is distributed in terms of position on the sky (images), wavelength or photon energy (spectroscopy), polarization, and time. Early examples of astronomical time series include observations of the positions of the stars and planets, the phases of Moon and planets, and sunspot numbers. The Wolf\footnote{This is sometimes called the `Wolfer sunspot number' in time series textbooks. See Izenman (1983).} sunspot data are a staple of introductory time series analysis books.

Most of the variable sources are effectively point-like on the sky, and most detectors are not sensitive to polarization, and so we will be concerned here with datasets that have at most two dimensions, characterising the brightness (often called {\it flux} by astronomers) against time and wavelength. Astronomers usually use the term {\it light curve} to mean estimates of the brightness of a target at different times, i.e. a univariate time series of the target's brightness, although other kinds of time series are also common in astronomy. Multivariate time series are generated by instruments that have some intrinsic energy resolution (ability to distinguish photons of different energy, or wavelength), or combining simultaneous data from several observatories that are sensitive in different wavebands (e.g. X-rays, optical and radio). 

At bottom, all of time series analysis is concerned with time varying signals that contain a non-trivial element of randomness. Depending on the target the astronomer may be interested in deterministic `signals' that are `contaminated' by random fluctuations. Such signals might be transients -- such as bursts of X-rays from the surface of a neutron star, of high-energy radiation from gamma-ray bursts (GRBs), or supernova -- or periodic signals such as from rotating stars, pulsars (radio pulsations due to neutron star rotation), or modulations due to the periodic orbits of binary stars, or planets. In these cases the astronomer is usually interested in recovering the deterministic component and testing models or estimating parameters, e.g. burst luminosity and decay time, rotation period, etc. In other cases the `noise' itself may represent the fundamentally stochastic output of an interesting physical system, as in turbulent accretion flows around black holes. Here the astronomer is interested in comparing the statistical properties of the observations to those of different physical models, or using the intrinsic luminosity variations to `map out' spatial structure. These projects, and many others, are completely dependent on time series data and analysis; our only access to the properties of physical interest is through their signature on the time variability of the light we receive.

Many of the different types of variable astronomical objects, and the types of variability they exhibit, are introduced in the nice review by Eric Feigelson (Feigelson 1997) given at a  conference on Time Series Analysis in Astronomy and Meteorology, that took place in 1993\footnote{Data files for all the examples in Feigelson's review are available from {\tt http://xweb.nrl.navy.mil/timeseries/timeseries.html}.}. In the two decades since that meeting the quality and quantity of astronomical time series has increased dramatically. In the field of X-ray astronomy, NASA's {\it Rossi X-ray Timing Explorer} produced thousands of datasets containing high frequency timing data (timescales as short as $\lsim$millisecond) from the Proportional Counter Array (PCA), and $15$ years of daily monitoring of the X-ray brightness of X-ray binaries with the All-Sky Monitor (ASM). The {\it Swift} mission has captured gamma-ray, X-ray and optical data of hundreds of GRBs and other high-energy transients. Optical astronomy has been transformed by large ground-based telescopes (the $8$m-class) and the now ubiquitous CCD detectors, and can routinely provide high-precision (in both absolute time and brightness) light curves of variable targets. Powerful optical time-domain surveys are now being undertaken by dedicated robotic telescopes in the search for timing signals of e.g. extra-solar planets, supernovae, trans-Neptunian objects, dark matter candidates, producing unprecedented quantities of time domain data. For example, the {\it CoRoT} and {\it Kepler} space missions (Auvergne et al. 2009; Borucki et al. 2010) and the MACHO, WASP, Pan-STARRS ground-based projects (e.g. Kaiser et al. 2002). Radio astronomy is currently enjoying something of a renaissance, and the Dutch-led {\it LOFAR} observatory will be sensitive to radio transients on timescales from milliseconds to months (Fender 2011). 

When writing an article on such a broad subject the most difficult question is not what to include, but what to leave out. What follows is rather parochial, being focussed on a just a few of the active areas and challenges that I have some direct experience with. In particular I will mostly focus on persistent variability, not transients. Although some of the most interesting astrophysical objects -- such as supernovae (Kulkarni 2011), pulsar flares (Bernardini 2011), gamma-ray bursts (GRBs; Gehrels et al. 2009), and stellar capture events (Burrows et al. 2011) -- are discovered through their transient emission, such sources are the exclusive subject of a another Royal Society discussion meeting this year (April 2012). Some of the other challenges in time domain astronomy that I have no room to discuss include high-precision pulsar timing (Liu et al. 2011), astroseismology (Appourchaux 2011) and planet detection (Gregory 2011; Ford et al. 2011).


\begin{figure}
  \centering
	\includegraphics[width=12.0cm, angle=0]{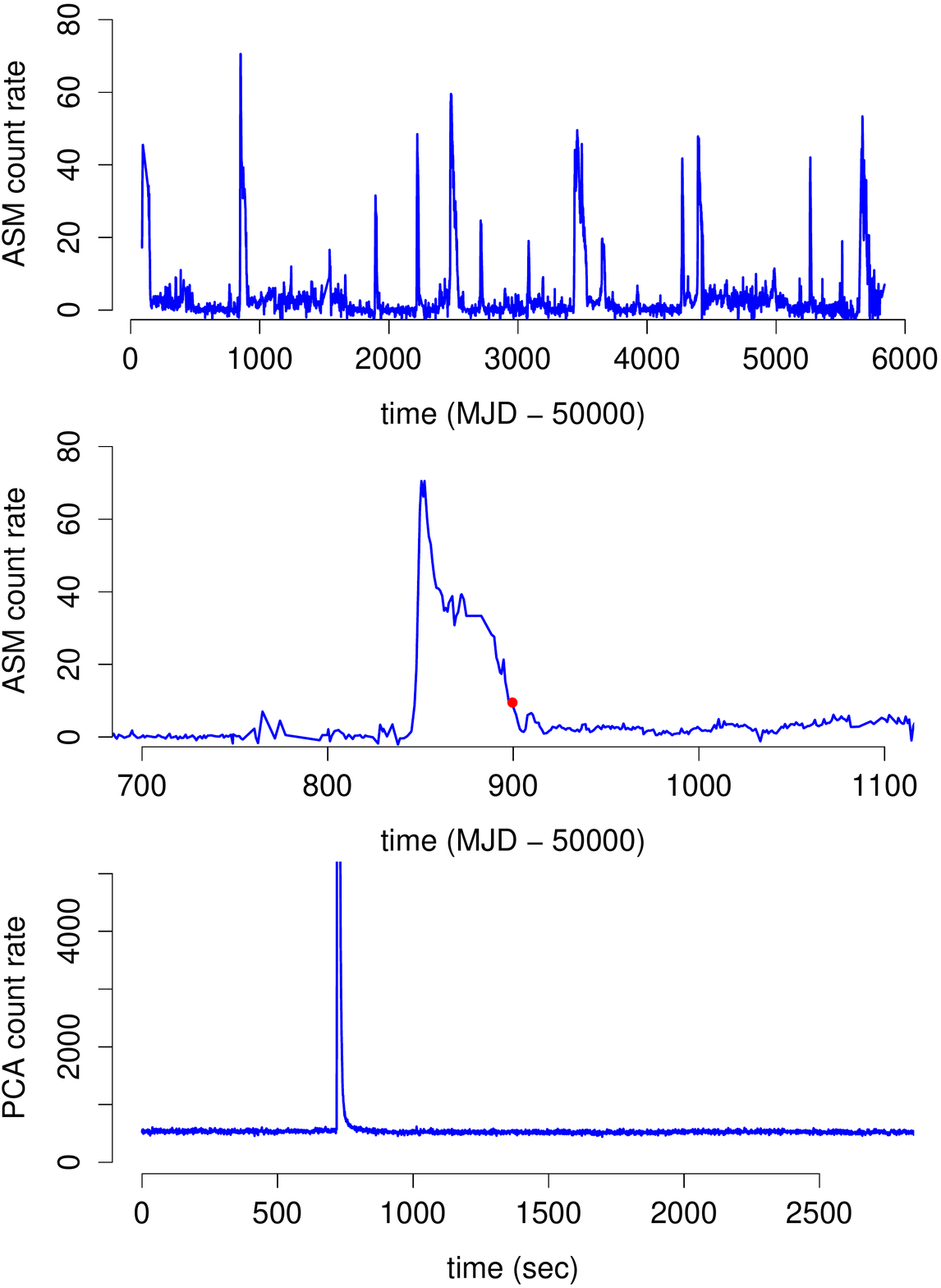} 
\caption{Examples of {\it slow} and {\it fast} variability. These show the X-ray brightness of the neutron star X-ray binary system 4U$1608-52$ as recorded by {\it RXTE}. (MJD$=$Modified Julian Date, one of the standard time systems astronomers use.) The upper panel shows the long-term evolution of the system (as observed by the {\it All Sky Monitor} instrument): the source goes through a series of outbursts lasting days--weeks separated by extended quiescent intervals. The middle panel emphasises one particular outburst from early 1998. The dot indicates the time of the observation shown in the bottom panel (27 March 1998), a single observation lasting $<$1 hour with the {\it Proportional Counter Array} (PCA) instrument, shown at a time resolution of $\Delta t = 1$s. On this time resolution we see two prominent features: relatively stable emission at a count rate of $\sim 500$ ct s$^{-1}$ and an X-ray burst (with a roughly exponential profile, $e$-folding time $\sim 5$s) with a peak count rate a factor $\sim 80$ above the previous level (the peak of the burst is not shown). The section of the data not affected by the burst shows small-scale variations on a range of timescales that are not apparent in this plot. 
}
\label{fig:timeseries}
\end{figure}


\section{Fast variability}
\label{sect:this}

Apart from a few rare exceptions (see below) most astronomical observations of particular targets are short, lasting minutes or hours. Observations with ground-based telescopes (e.g. in the radio, infra-red or optical bands) are limited by e.g. the length of a night, or the visibility of the target on the sky. Earth orbiting telescopes may have their observations interrupted as the Earth periodically blocks the telescope's view of the target, or due to gaps in the telemetry. For both types of observatory the observations are limited by other scheduling and operational constraints. As a result astronomers tend to treat within-observation and between-observation variability slightly differently. Variability within a single, continuous observation is usually limited to timescales $\lsim$few hours. I shall refer to this within-observation variability as {\it fast} variability. Time series covering much longer timescales can be built up through repeat observations of the same target, often through dedicated {\it monitoring campaigns}, and can in principle access variations on timescales of $\sim$few hours--decades but with generally irregular time sampling. Below I shall refer to such inter-observation variability as {\it slow} variability.

\begin{figure}
  \centering
	\includegraphics[width=13.0cm, angle=0]{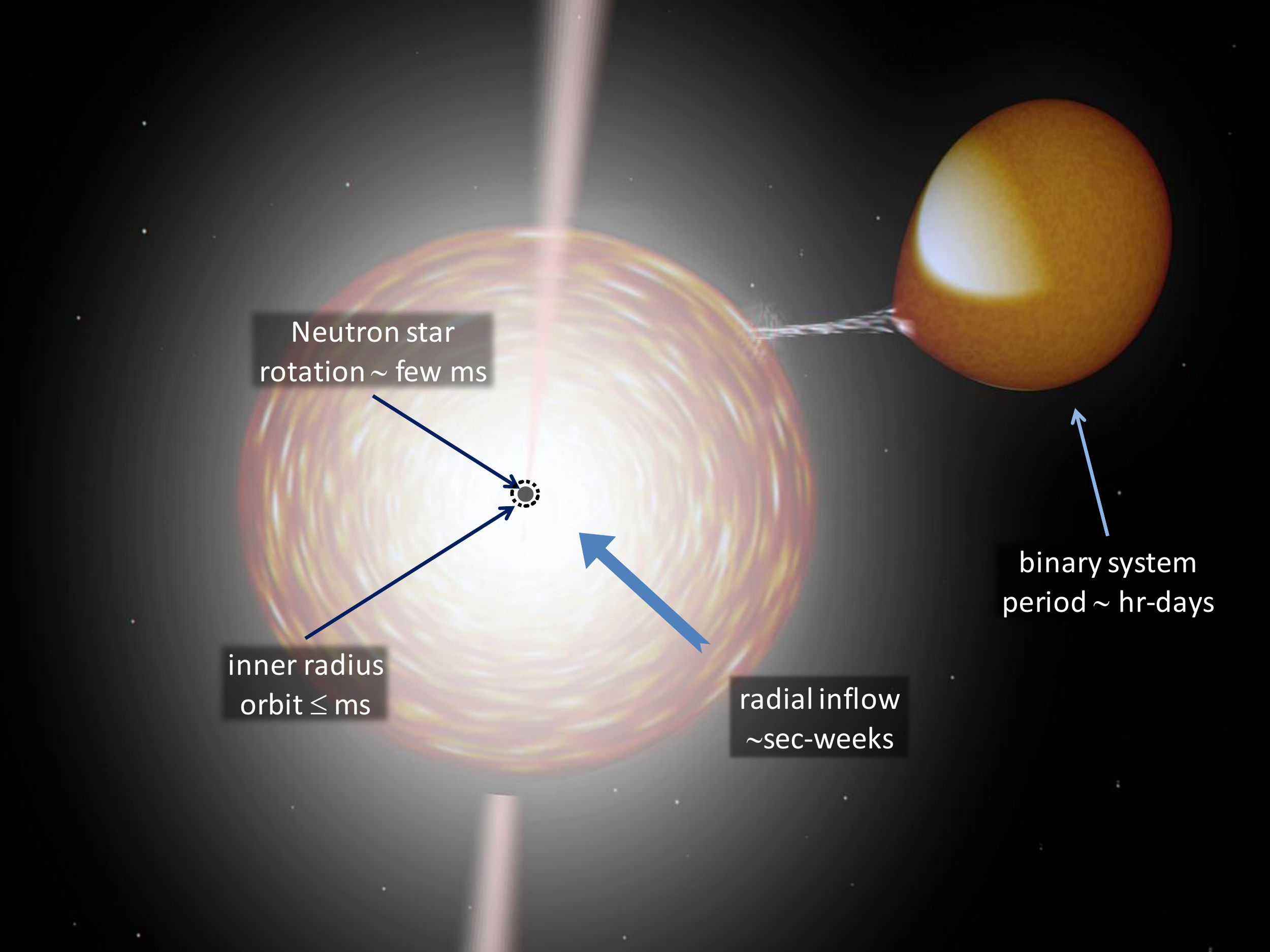} 
\caption{Illustration of an X-ray binary system comprising a compact object (a neutron star, in some systems a black hole) accreting matter lost from a (normal) companion star via an accretion disc. The accretion disc is expected to radiate from the infra-red through X-rays. Other processes (e.g. a relativistic jet) may provide another source of radiation from radio though hard X-rays/$\gamma$-rays. All these may be modulated by variations in the turbulent accretion flow, and its interaction with the compact object, over a very wide range of timescales. (Image generated using Rob Hynes {\tt binsim} code.)
}
\label{fig:binary}
\end{figure}

I will discuss some examples of fast variability from the field of X-ray astronomy. But first, I should explain what kind of data X-ray astronomy deals with.
Data from observations of particular targets by orbiting X-ray detectors are usually packaged as {\it event lists}, essentially just tables detailing the properties of detector events (most, but not all, of which are triggered by genuine X-ray photons). Each event is associated with several properties that give the arrival time, energy channel and position on the detector, as well as other information. If we are interested in the variability of the target we form a light curve by selecting a subset of the events\footnote{For example, events that fall within particular energy channels and within a particular part of the detector image. Not all events recorded by the detector will be caused by X-rays from the target source. Some event selection is required to reduce or remove background events, non-X-ray events, and other contaminants.} and making a histogram of the event times. See the excellent book by Arnaud et al. (2011) for more on X-ray detectors, data, and processing. 

\begin{figure}
  \centering
	\includegraphics[width=13.0cm, angle=0]{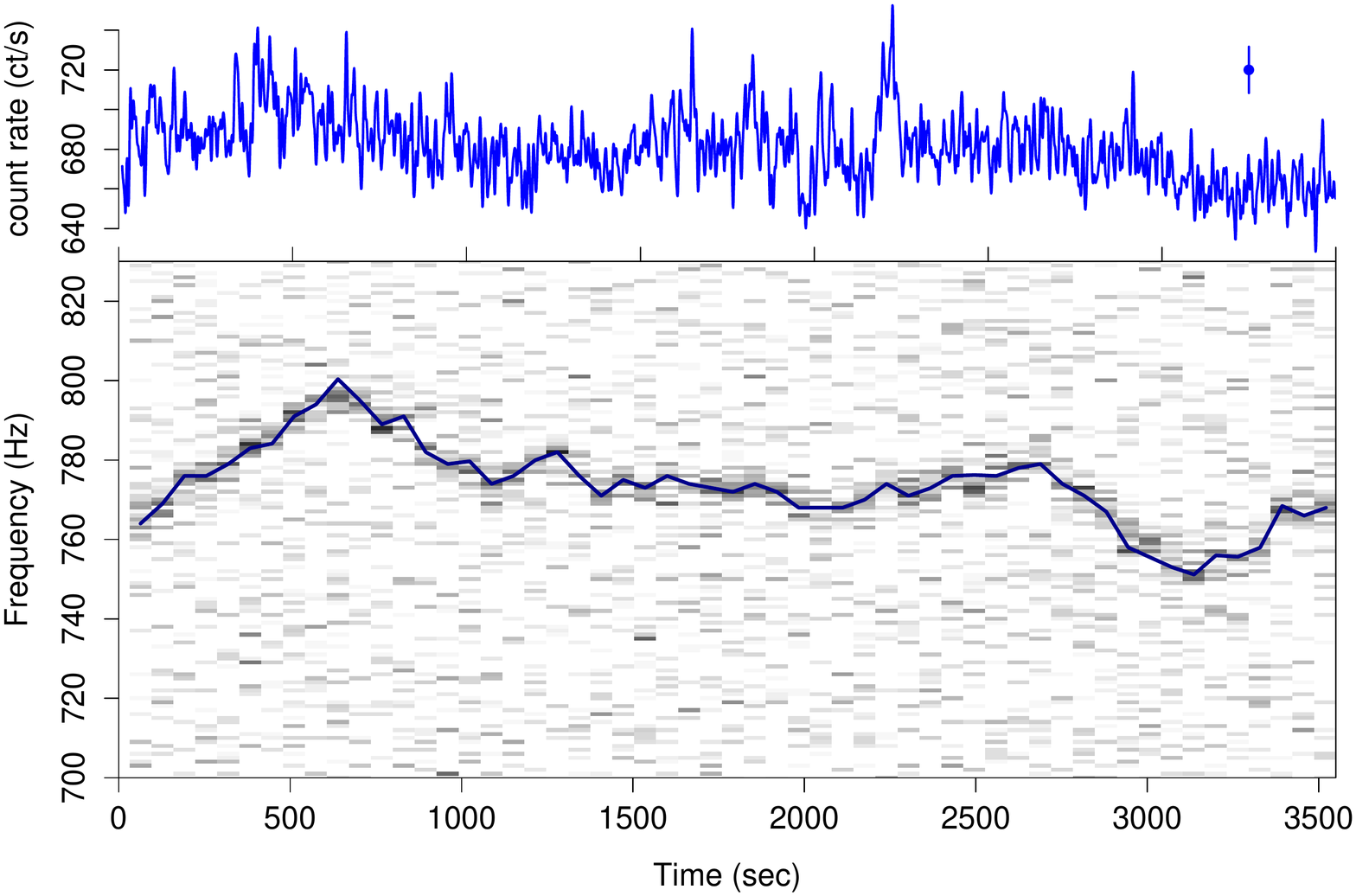} 
\caption{Upper panel: time series of 4U$1608-52$, as observed by {\it RXTE} on 27 March 1998 for about an hour. The data are shown averaged in $5$ sec time bins. The size of a typical error bar is given in the top right corner. Lower panel: dynamic (short-time) power spectrum estimates. Each vertical slice is a section of the power spectrum estimated by averaging $64$ periodograms each computed from $1$ second of data sampled at $2048$ Hz. These reveal an excess of power at $\sim 780$ Hz; the frequency of this peak is itself variable on short timescales. The solid curve shows the maximum likelihood estimate (MLE) of the peak frequency after fitting a simple model to each of the power spectral estimates (see Barret \& Vaughan 2012).}
\label{fig:dps}
\end{figure}

Figure \ref{fig:timeseries} shows example light curves for a bright, variable X-ray source, 4U$1608-52$, based on observations by the {\it Rossi X-ray Timing Explorer} ({\it RXTE}). The source is an {\it X-ray binary}, a type of binary star system in our Galaxy comprising a low-mass `donor' star transferring matter onto a neutron star via an disc-shaped accretion flow (Fig. \ref{fig:binary}). The top panels show the {\it slow} variability recorded by the {\it All-Sky Monitor} (ASM) instrument, shown as $\approx$daily averages. The bottom panel shows the light curve of a single observation lasting $<$1 hour with the {\it Proportional Counter Array} (PCA) instrument. The PCA data were recorded with a time resolution of $2^{-16}$ s but are shown `binned' to a resolution of $\Delta t = 1$ s. The ASM data show accretion {\it outbursts}, where the source becomes X-ray bright, reaching luminosities of $\gsim 10^{30}$ W ($\gsim 4 \times 10^3$ solar luminosities), typically lasting days--weeks (top two panels). Outbursts are thought to be driven by large-scale instabilities in the accretion flow (Lasota 2001). During outbursts the source shows both persistent and transient variability. The bottom panel of  Figure \ref{fig:dps} illustrates a fast transient phenomenon, an {\it X-ray burst} lasting about a minute caused by runaway thermonuclear `burning' on the surface of the neutron star (see Strohmayer \& Bildsten 2006 for a review). 

\begin{figure}
  \centering
	\includegraphics[width=12.0cm, angle=0]{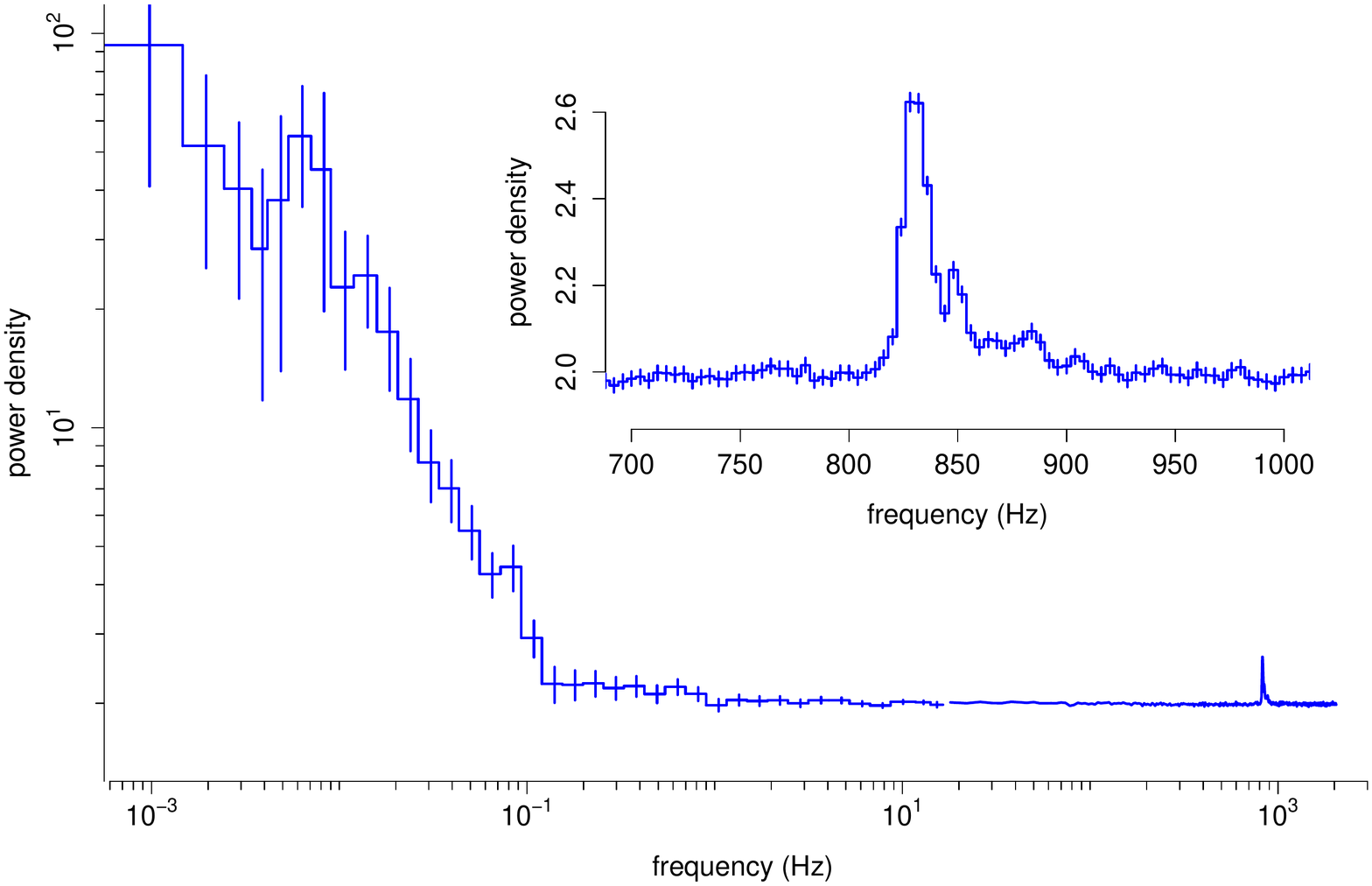} 
\caption{Power spectrum estimated using three observations taken on 03 March 1996 spanning six decades in frequency. At low frequencies we see the broad-band noise spectrum of the source. Above $\sim 1$ Hz the spectrum flattens to a level $\approx 2$ due to Poisson noise (i.e. the random arrival time of X-ray events even for a source of constant intrinsic brightness). At frequencies close to $1$ kHz there is a narrow spectral feature, known as a kHz QPO (shown inset).
}
\label{fig:broad_psd}
\end{figure}


\subsection{The power spectrum of fast variations}

Figure \ref{fig:timeseries} clearly shows a wide range of variability types. Here we will concentrate on the persistent, fast variability. This may not be so obvious from Figure \ref{fig:timeseries} because of the choice of time binning and the appearance of the X-ray burst. 
Figure \ref{fig:broad_psd} shows a power spectrum taken from a series of observations of the same source during which the source was bright and there were no X-ray bursts. As in many areas of the physical sciences, the power spectrum is often used by astronomers to characterise the variations in more detail. It describes the distribution of {\it power} (squared variability amplitude) as a function of temporal frequency ($\sim 1/$timescale). 

In the case of the 4U$1608-52$ data, the power spectral estimate shows some weak low frequency power, a flat spectrum $\gsim 1$ Hz caused by Poissonian fluctuations in the X-ray count rate, and a narrow bump at $\gsim 700$ Hz. This excess of power at high frequencies is a {\it quasi-periodic oscillation} (QPO), a concentration of variability power in a narrow range of frequencies. This is distinct from `strictly periodic' variability (repeating perfectly after a fixed interval $T$) that would appear as unresolved narrow peaks in the spectrum, and `noise' where power is spread over a wide range\footnote{Astronomers typically define QPOs in terms of the width of the peak, $\delta f$, relative to its centroid frequency, $f_0$: $Q = f_0 / \delta f$ the {\it quality factor} (also called {\it coherence}) of the oscillation. Periodic, quasi-periodic and noise processes are loosely defined in terms of their $Q$ values: unresolved peaks with very high $Q$ are strictly periodic, peaks with finite $Q$ down to $Q=2$ are defined as QPOs (quasi-periodic), and even broader  features ($Q < 2$) are usually called noise (aperiodic).} The peak in the 4U$1608-52$ power spectrum is an example of a ``kHz QPO'' (because their frequencies range to $\gsim 10^3$ Hz in some cases) and have been detected in $\sim 20$ different neutron star X-ray binaries (van der Klis 2006, 2007). They are of great interest to X-ray astronomers, being the fastest (persistent) variations of any known astronomical source, and thought to be a result of the relativistic gravity acting on the inner accretion disc close to the neutron star surface (Stella \& Vietri 1999) since the orbital period close to the neutron star surface is $\sim 1$ ms.


\subsection{Estimating the power spectrum}

How do astronomers estimate the power spectrum from these data? The standard procedure is to follow the Bartlett method (Bartlett 1948):
\begin{itemize}
\item from the event list form an evenly sampled time series $x_t$ with a bin time size $\Delta t$
\item break this into $M$ non-overlapping intervals of length $N$
\item compute the periodogram of each interval
\item average the $M$ periodograms. 
\end{itemize}
If the (binned) light curve is $x_t$, and $X_j$ is its (complex valued) Discrete Fourier Transform (DFT) at Fourier frequency $f_j$, then the periodogram is $A|X_j|^2 = A X_j^{\ast} X_j$ where $A$ is a normalising term (Vaughan et al. 2003) and $X^{\ast}$ is the complex-conjugate of $X$. Under simplifying assumptions (e.g. that the data are a realisation of a linear, stationary and ergodic process) the Bartlett estimator should converge to the true power spectrum $S(f)$ in the limit of large $N$ and $M$. See van der Klis (1989) for discussion of standard power spectral methods as applied in X-ray astronomy. This has become the standard method of power spectral estimation in astronomy when the data are evenly sampled (below we mention the case of uneven time sampling). The result is a power spectral estimate at each of the $N/2$ Fourier frequencies $f_j = j / N \Delta t$, which should have a Normal distribution (for large $M$) with standard deviation $S(f_j) / \sqrt{M}$.

This is completely conventional analysis and has been popular in astronomy for decades. However, researchers working on time series in fields such as biology, geology, medicine, engineering, etc.  often use different methods, e.g.: Tukey--Blackmann (TB), auto-correlation function (ACF), partial ACF, Maximum Entropy/Burg method, ARMA modelling, and so on. To a large extent these are all just slightly different ways of calculating and estimating the same underlying function, the power spectrum. But the Bartlett/periodogram approach has a number of features that help maintain its popularity with astronomers. These include: the sensitivity of the periodogram to nearly sinusoidal variations, as might be expected from the rotation of stars or the orbits of binary stars, the ease with with the DFT can be computed (via Fast Fourier Transform, FFT, algorithms), and its statistical properties. The Bartlett method yields power spectrum estimates at each frequency that are (asymptotically, for larger $M$ and $N$) independent of each other and Normally distributed, which permits the use of simple least squares fitting of parametric models (for maximum likelihood estimation of parameters, and goodness-of-fit tests). 


\subsection{Tracking the evolution of the power spectrum}

The count rate of the source during the observation shown in Figure \ref{fig:dps} was $\sim 680$ counts/s, meaning we have $\sim 1$ X-ray count per characteristic timescale of the kHz QPO, and the QPO is clearly broad albeit with a high $Q \sim 50$. This means we cannot easily examine the kHz QPO in the time domain: there are too few counts/cycle to recover the waveform of the oscillation, and if we average (`fold') the light curve on the period corresponding to the peak frequency in the power spectrum -- in order to make up for the lack of counts per cycle -- the result carries very little meaning because the QPO is not a coherent oscillation with a single, strictly repeating period. Future X-ray missions with larger collecting area may be capable of resolving these $\sim$ms variations in the time domain (e.g. Feroci et al. 2011).

\begin{figure}
  \centering
	\includegraphics[width=13.0cm, angle=0]{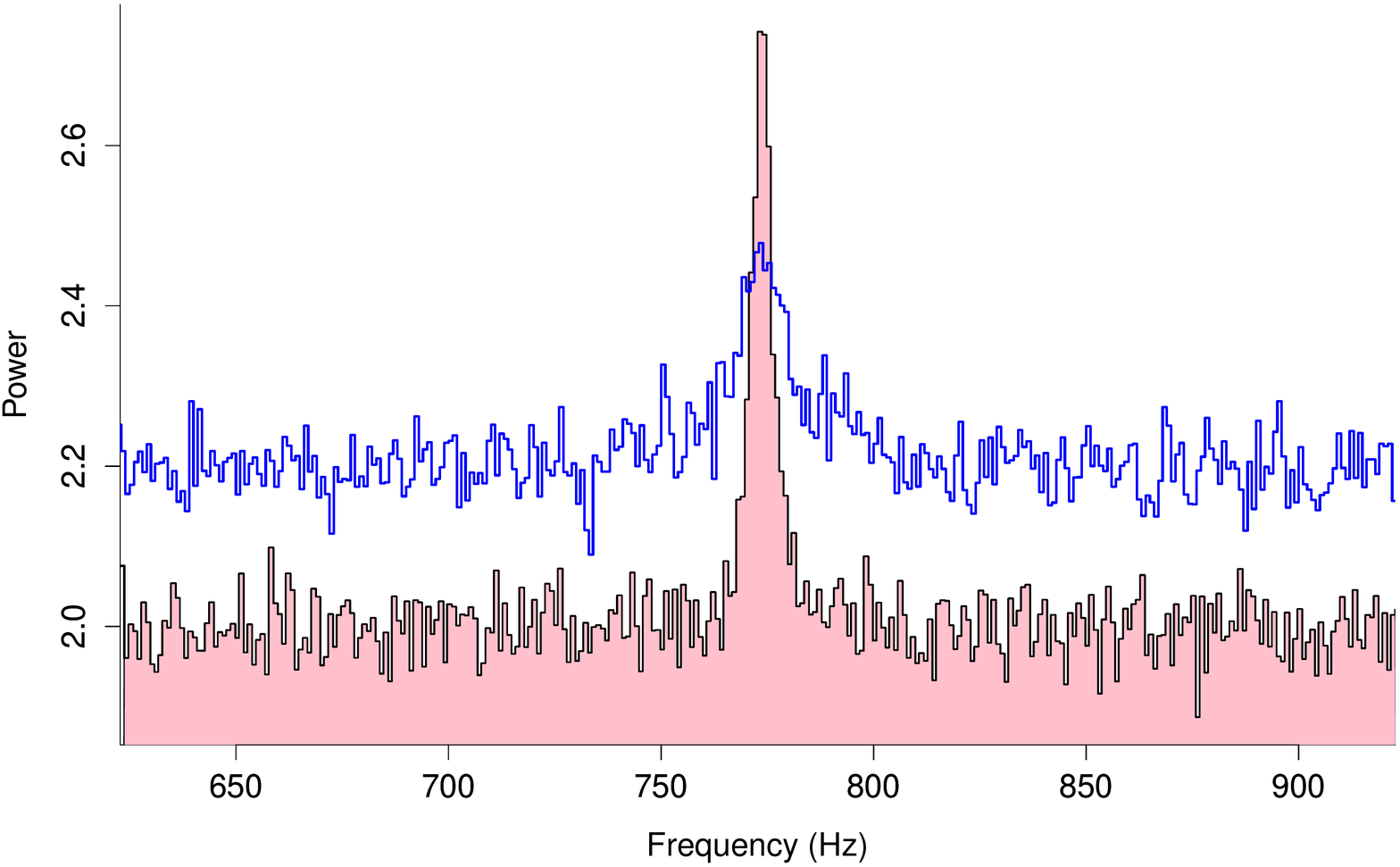} 
\caption{The two histograms are different estimates of the power spectrum of 4U$1608-52$ averaged over the entire observation shown in Figure~\ref{fig:dps}. In both cases we see a mixed spectrum comprising a flat spectrum due to Poisson noise in the photon counting detector, and a concentration of power in the form of a QPO. The upper curve, which has been shifted upwards by $0.2$ for clarity, is the simple spectral estimate obtained by averaging the periodograms of each $1$ sec of data. The lower curve is the average obtained after shifting the frequency axis of each periodogram such that the peak is aligned with a common frequency. This `shift-and-add' average reveals a much narrower and better-defined spectral peak.}
\label{fig:sum_dps}
\end{figure}

The width (incoherence) of the QPO is partly explained when we examine the dynamic power spectrum, as shown in the lower panel of Figure~\ref{fig:dps}. This shows the variability power as a function of frequency and time, and clearly reveals the QPO frequency to be variable on short timescales. But again the data impose limits on what we can observe. It is necessary to average over enough data (e.g. $\sim 50 \times 1$ sec periodograms) to even detect the QPO, this sets a limit on the minimum timescale on which we can observe variations in the QPO properties. Nevertheless, we can go some way to removing the effect of {\it frequency drift} on the width of the QPO by using the `shift-and-add' method to align the short-time power spectrum estimates to a common frequency (Mendez et al. 1998). Figure \ref{fig:sum_dps} shows the power spectrum estimated by averaging the $1$ sec periodograms over the entire observation with and without the frequency-shift correction. In this case the frequencies were estimated by fitting the short-time periodograms with a simple parametric model using maximum likelihood (ML) fitting as discussed in Barret \& Vaughan (2012). The `shift-and-add' method does indeed recover a narrower QPO, although not without bias.  

In Barret \& Vaughan (2012) we produced a time series of QPO frequencies, and then examined the power spectrum of this by fitting models using maximum likelihood. This is in essence a {\it hierarchical} or {\it multi-level} analysis: we model the power spectrum of frequency variations of a QPO, which in turn are estimates derived from modelling short-time power spectral estimates of the light curve, which in turn is based on an even higher time resolution event list. Based on the power spectrum of the QPO frequency variations we can estimate broadening of the QPO due to frequency drift on any given timescale, and correct this to recover the unobserved `intrinsic' width of the QPO.  Using this method the recovered $Q$-value is at least as high as $\sim 260$. Such high coherence is a challenge for most models of the behaviour of matter close to the neutron star (e.g. Barret et al. 2005). The power spectrum of frequency variations is limited at high frequencies by the integration time required to determine the QPO frequency, and at low frequencies by the length of the observations. ({\it RXTE} is in a low Earth orbit which means observations are interrupted at least every $96$ mins.)

This particular example was used to illustrate some of the complexities that arise when analysing even quite strong timing signals. In this case the source is bright and the variations of interest (the kHz QPO) are relatively strong, persistent, and sampled for a large number of cycles (and therefore well resolved in Fourier space). And yet recovering its basic parameters remains a challenge. Issues of high frequency timing like this have been a concern in X-ray astronomy, and to some extent radio astronomy (e.g. pulsar timing), for many years. High time-resolution instruments are also used in optical and infra-red (IR) astronomy (e.g. Marsh 2008), so these kinds of problems are certainly not restricted to X-ray observations.


\section{Non-linear time series}

Most work on fast timing of persistently variable astronomical sources is based on power spectral methods and related second-order statistics (cross-spectrum, auto-correlation function, etc.). These contain information on the statistical moments up to second order only. If the variability processes are linear, Gaussian and stationary then no information is lost, everything is contained in the second order statistics. (The phases of the Fourier components are random, and so it is no real loss to throw them away when we square the Fourier transform to estimate the power spectrum.) This remains an implicit (and often untested) assumption, rather than than an open area of investigation, which may in part be due to the fact that constructing physical models of the power spectra has proved to be a serious challenge. However, the assumption of linearity and Gaussianity cannot be strictly true since we observe high amplitude fluctuations (e.g. rms $\sim 20$ per cent), but the luminosity fluctuations cannot extend below zero. Other statistics, that are sensitive to the non-Gaussian and non-linear properties of data, may provide more discriminating tests of existing models or point the way towards new ideas.

One potentially powerful diagnostic is the bispectrum, or bicoherence. Where the power spectrum is a second-order spectrum, which can be thought of in terms of the product of two Fourier transforms [$X(f_j)$ $X^{\ast}(f_j)$], the bispectrum is a third-order spectrum formed from the product of three Fourier transforms [$X(f_j)$ $X(f_k)$ $X^{\ast}(f_j + f_k)$]. The computation and interpretation of the bispectrum (and higher-order spectra in general) is treated by e.g. Brillinger \& Rosenblatt (1967) and Nikias \& Petropulu (1993), and it has been applied in geology (Rial \& Anaclerio 2000), plasma physics (Kim \& Powers 1979) and speech processing (Fackrell \& McLaughlin 1996).

The bispectrum is a complex quantity defined for the pair of frequencies $(f_j,f_k)$.  The phase of the bispectrum is often called the \emph{biphase} and its modulus-squared amplitude is called the \emph{bicoherence} (once it has been suitably normalised).  The bispectrum provides a measure of the nonlinear interaction of frequency components in the time series. If the variations at the frequencies $f_j, f_k, f_j+f_k$ are independent of each other they will have independent, random phases, and the bispectrum will tend to zero. If the variations at frequencies $f_j$ and $f_k$ are multiplicatively coupled they will contribute a `side-band' component at $f_j+f_k$. The phases at the three frequencies will be correlated, and the bispectrum will not average to zero. These properties make the bispectrum sensitive to the non-Gaussian and non-linear characteristics of time series. 

\begin{figure}
  \centering
  \hbox{
    \hspace{-1 cm}
  	\includegraphics[width=8.5cm, angle=0]{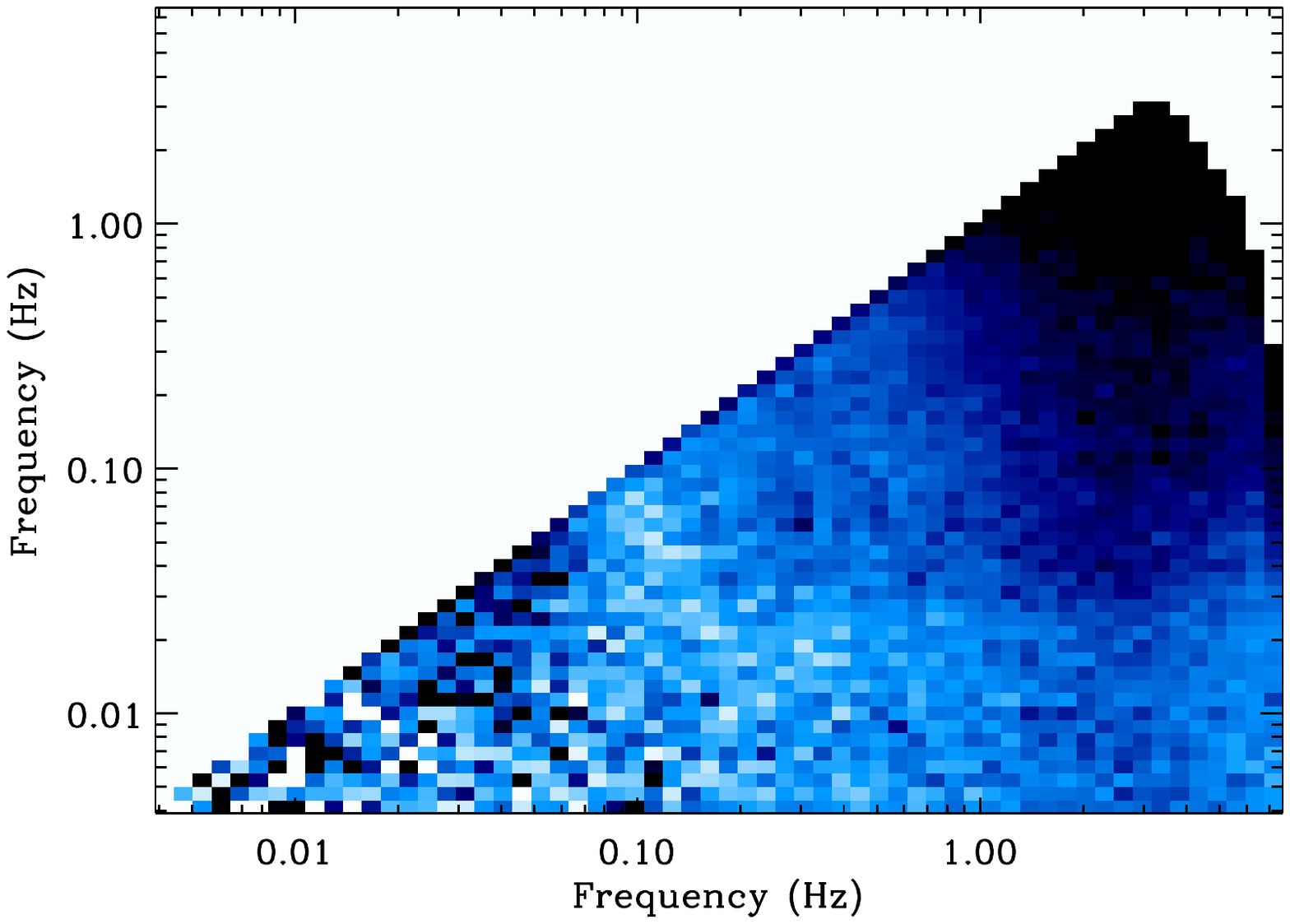} 
    \hspace{-2 cm}
	  \includegraphics[width=8.5cm, angle=0]{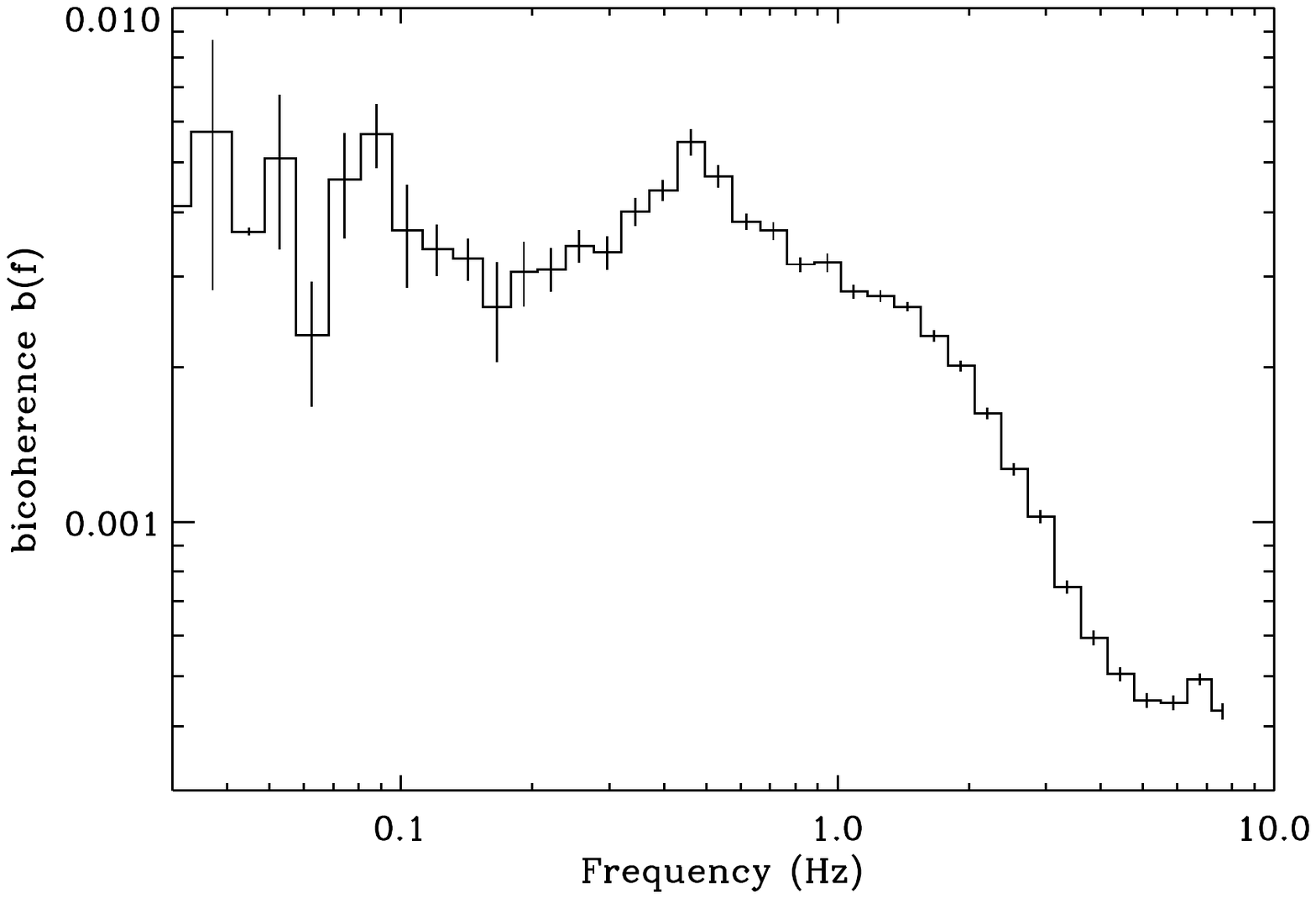} 
	  }
  \vspace{-5 cm}
\caption{Bicoherence from X-ray observations of Cygnus X-1 data. The left panel shows the bicoherence from the `principal domain' of bi-frequencies $f_j,f_k$ (the principal domain is defined as those pairs of frequencies for which $f_j$ spans the full range of Fourier frequencies $[0,1/2\Delta t]$, with the additional constraints $f_k \le f_j$ and $f_j + f_k \le 1/2\Delta t$). The effect of detector noise has been subtracted (to first order). The bicoherences were binned in equal logarithmic frequency intervals along each axis, and colour coded such that lighter shades correspond to higher bicoherences. In order to better illustrate the structure in the bicoherence the right panel shows the bicoherence compressed into one dimension as a function of $f_j + f_k$ (by summing over lines of constant $f_j+f_k$). The error bars come from the scatter of individual bicoherences within each logarithmic frequency bin (as individual bicoherence estimates are not independent at different bifrequencies, these should be taken as illustrative only). From Vaughan \& Uttley (2007).
}
\label{fig:bicoh}
\end{figure}

Figure \ref{fig:bicoh} shows the bicoherence of the black hole X-ray binary system Cygnus X-1, estimated using a long series of X-ray observations taken in 1996 with {\it RXTE} (Vaughan \& Uttley 2007). (The source shows strong, persistent and reasonable stationary aperiodic variability, with not strong QPOs.) The non-trivial structure in the bicoherence indicates frequency coupling, but interpreting the bicoherence is not straightforward. The bicoherences at each bifrequency ($f_j$ , $f_k$) are not independently distributed, and the Poisson noise in these data causes spurious bicoherence at high frequencies (the Poisson noise fluctuations are uncorrelated with the source flux but are not independent, since the amplitude of Poisson fluctuations scales with $\sqrt{flux}$). At high frequencies the Poisson noise dominates over the intrinsic variations in the source flux (which drop off rapidly with increasing frequency). Nevertheless, bicoherence is now starting to be used as a powerful test for more advanced models of variability from black holes and neutron stars (e.g. Maccarone \& Coppi 2002; Maccarone et al. 2011).

\begin{figure}
  \centering
  \hbox{
    \hspace{-1 cm}
  	\includegraphics[width=8.5cm, angle=0]{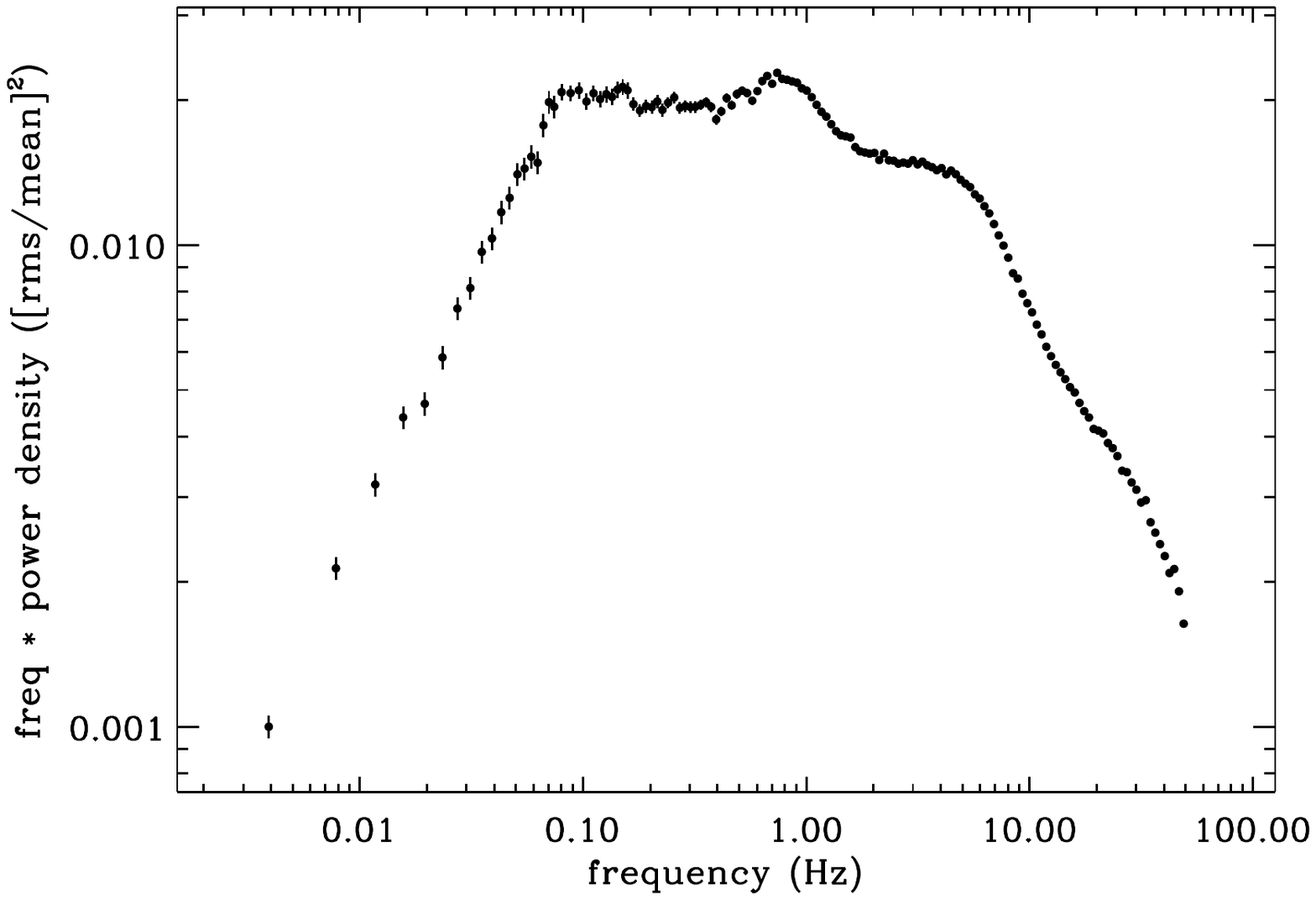} 
    \hspace{-2 cm}
	  \includegraphics[width=8.5cm, angle=0]{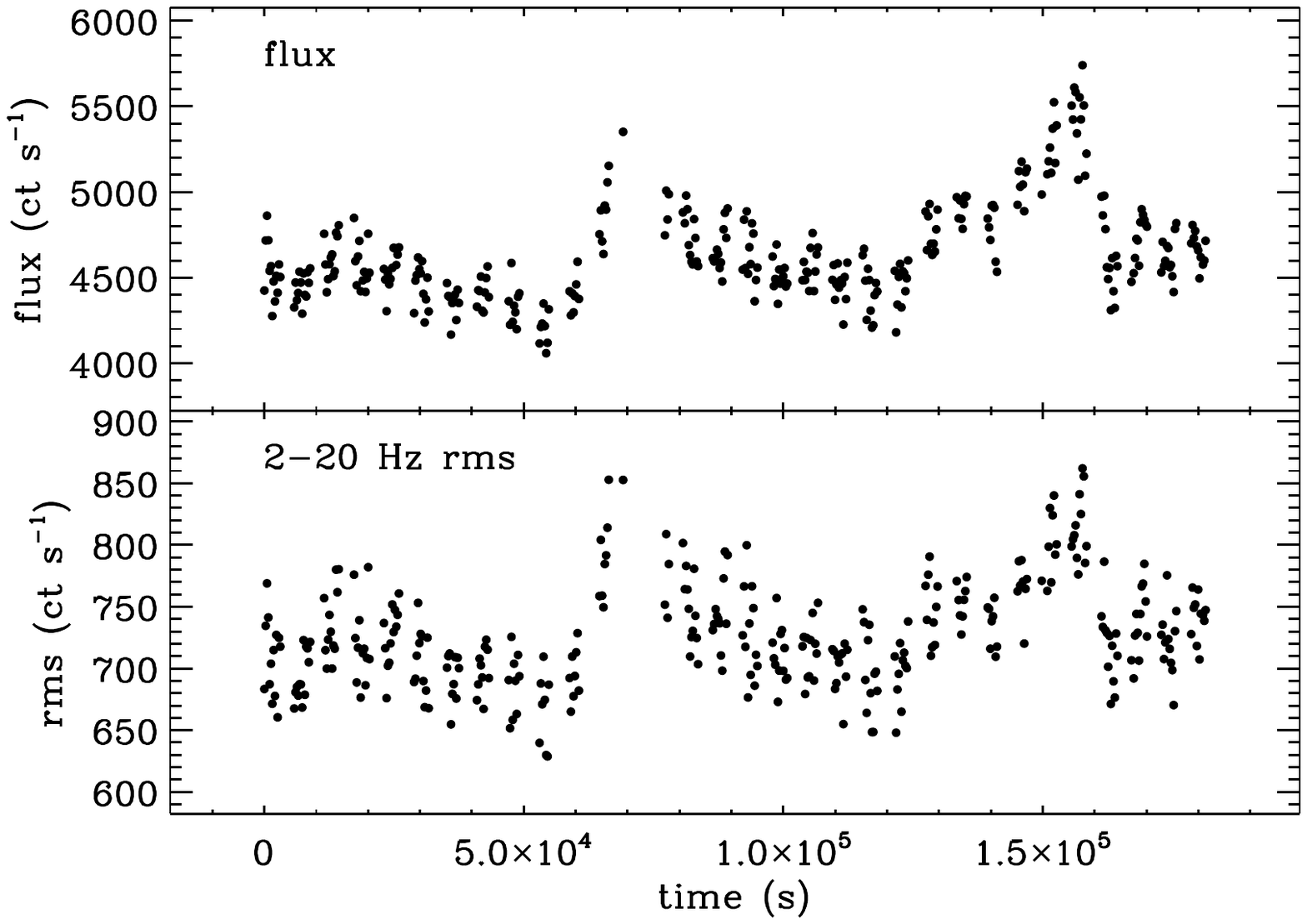} 
	  }
  \vspace{-5 cm}
\caption{Variability properties of the black hole X-ray binary Cygnus X-1. 
The data are from an {\it RXTE} observation taken on 16-19 Dec 1996, sampled in bins of $\Delta t = 2 \times 10^{-8}$ s (3.90625 ms). 
{\it Left}: 
the power spectrum plotted as frequency$\times$power on a log-log scale, which emphasizes which frequencies dominate the variance (integrated spectrum). In this $f \times S(f)$ representation a flat spectrum means equal power per logarithmic frequency interval (i.e. $f^{-1}$ spectrum). This shows a broad, noise spectrum that roughly resembles a broken power law spectral shape, with a $f^0$ spectrum below $\sim 0.06$ Hz, a roughly $f^{-1}$ spectrum (which appears flat in the $f \times S(f)$ representation) in the range $\sim 0.06 - 6$ Hz, and a $f^{-2}$ spectrum at higher frequencies. 
{\it Right}: Time series of the mean flux (upper panel) from $256$ s averages of the X-ray count rate, and the $2 - 20$ Hz rms calculated from the integral of the periodogram of each $256$ s interval (after subtracting the power due to Poisson noise). Clearly the rms and the flux are correlated. The gaps in the coverage are due to observational constraints. From Vaughan \& Uttley (2007). See also Uttley, M$^{c}$Hardy \& Vaughan (2005).
}
\label{fig:cyg}
\end{figure}

Figure \ref{fig:cyg} illustrates a different approach to the same data, discussed by Uttley, M$^{c}$Hardy \& Vaughan (2005) and Vaughan \& Uttley (2007). During this particular observation the shape of the power spectrum of Cygnus X-1, at least at frequencies $\sim 10^{-3} - 10^2$ Hz appears to remain quite constant through more than two days of observation. However, its amplitude does not. The normalisation but not the shape of the power spectrum seems to scale (on average) with the X-ray count rate: brighter intervals are also more variable intervals. This `rms-flux' relation is now quite well established in accreting systems (e.g. Heil, Vaughan \& Uttley 2012; Scaringi et al. 2012 and references therein). This can be thought of as an example of non-stationarity, since the variability properties are changing with time, or as a `static' non-linear transformation of the series\footnote{Strictly, we should consider non-stationary and non-linearity to be properties of models or processes, not data, which are but realisations of the process. In the present case the model may be described as either non-linear or non-stationary.}, since normalising an interval by its mean flux stabilises the variance (there is connection here to Box-Cox transformations). However, even after accounting for this effect there remain some residual non-linearities that are unexplained.


\section{Non-stationary time series}


\begin{figure}
  \centering
	\includegraphics[width=13.0cm, angle=0]{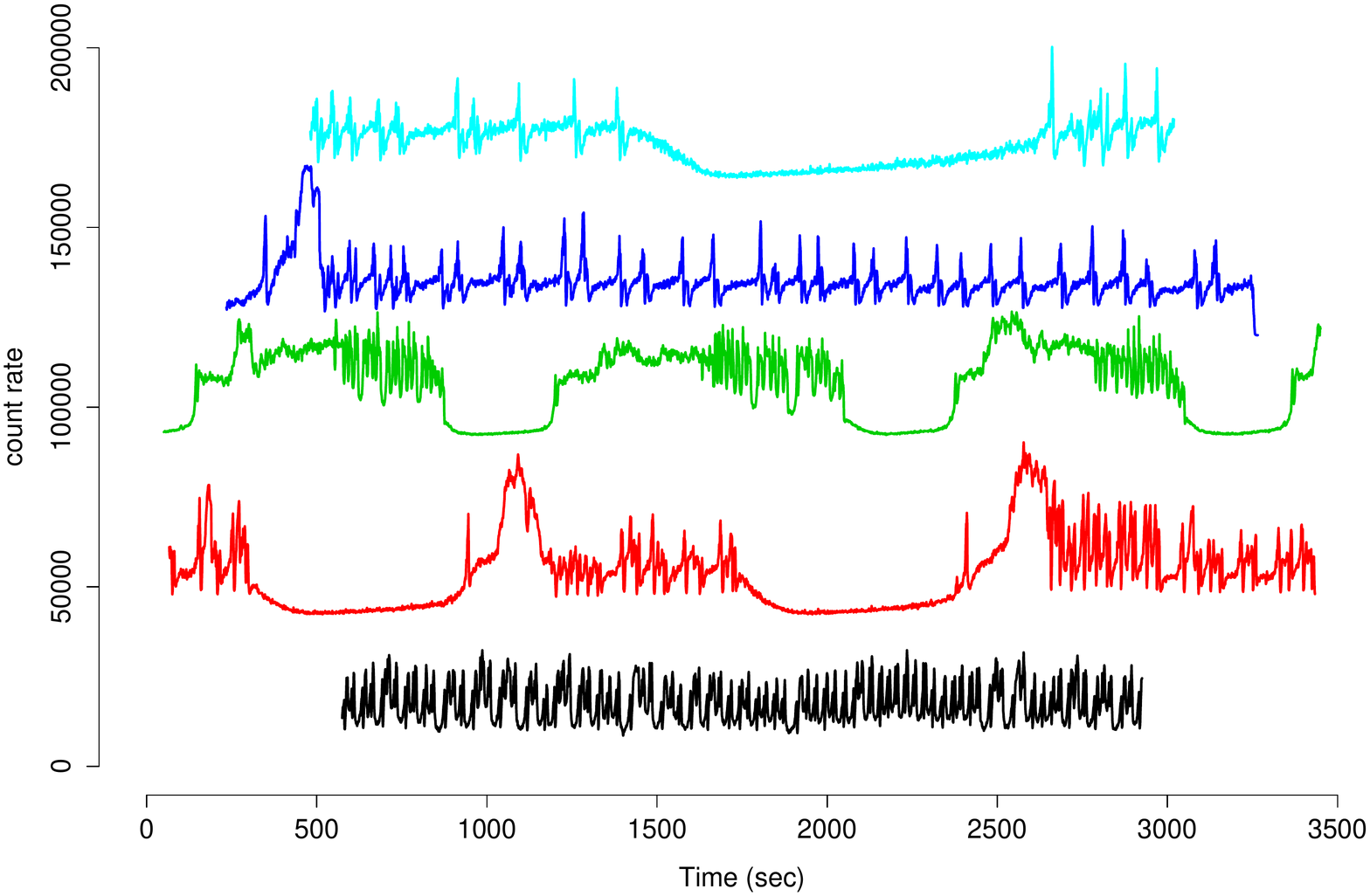} 
\caption{Examples of the curious patterns of variability exhibited by the luminous black hole X-ray binary GRS $1915+105$. The figure shows light curves with $\Delta t = 1$ sec from five different X-ray observations, shifted vertically for clarity. Each observation typically lasts $\lsim$1 hr. The same kinds of patterns can be found in observations of the same source separated by months or years and are accompanied by specific patterns of variability in other wavebands (Fender \& Belloni 2004). }
\label{fig:grs1915}
\end{figure}

We have so far simply assumed that the time series we are interested in are stationary (or, strictly, that they are random realisations of a stationary stochastic process). This is a necessary assumption for some analysis methods, e.g. to obtain a consistent power spectral estimate using the Bartlett method we assume that periodograms calculated for different time intervals are drawn from the same underlying process and so may be averaged to produce a meaningful estimate. But one thing we do know about these systems is that often the variability is not stationary. 

The kHz QPO shown in Figure \ref{fig:dps} is clearly not stationary because its frequency varies on timescales as short as $\lsim 100$ sec. However, it may be the case that this higher level of variability is itself stationary, i.e. the frequency variations can be considered as a stationary stochastic process (as with the rms-flux effect mentioned above). Even this is not true on longer timescales because these sources undergo periods of outbursts and quiescence (Figure \ref{fig:timeseries}), and move through a variety of `states' that display characteristically different variability (and other) properties. Figure \ref{fig:grs1915} shows some X-ray light curves of a black hole X-ray binary, GRS $1915+105$. This particular source seems to move through a few reasonably well defined `states' (with specific high frequency timing and energy spectral properties), but the pattern and speed of the transitions are quite remarkable, and remains largely unexplained (see e.g. Fender \& Belloni 2004).

This non-stationarity imposes another limit on what we can achieve with conventional methods. It may be the case that on short timescales the system is ``locally'' stationary, and so we can measure e.g. power spectral properties and examine their time evolution on the frequency-time plane. There is now a large volume of work on the evolution of power spectral features in X-ray binaries (see e.g. van der Klis 2006 for a review). But in some cases it may not be meaningful to estimate the power spectrum if the system is non-stationary on the shortest timescales on which we are able to estimate a power spectrum.


\section{Slow, persistent variability}

There are two facts about the fast timing of sources like X-ray binaries that help keep the field lively. The first is that there is a rich phenomenology: a range of power spectral features (noise, low frequency QPOs, kHz QPOs, pulsations) that evolve systematically with time, in addition to transient events (rapid bursts, accretion outbursts) that can be detected, identified, classified and compared to other observables in the hope of finding associations that might help explain their origins (Belloni 2007; van der Klis 2006, 2007). The second reason is more pragmatic: there is a lot of good data. There are thousands of observations from {\it RXTE} of X-ray binaries over a wide range of luminosities and states, covering frequencies up to kHz, often with high count rates, and there are now many good fast optical and infra-red time series, sometimes simultaneous with the X-ray ones. The high photon event rates (meaning a relatively small contribution to the power from Poissonian fluctuations), high time resolution and multiple observations means that in many cases we can compute very good power spectra (or related statistical summaries of the variability), in the sense of having small variance and low bias on the estimates. And where there is non-stationarity the data are often sufficient to resolve the varying power spectrum in the time domain. But not all astronomical time series are like this.

At the other end of the frequency spectrum we see much slower variations from active galactic nuclei (AGN). AGN are the luminous cores of galaxies thought to be powered by accretion onto a supermassive black hole ($M_{BH} \sim 10^6 - 10^9$ solar masses). Over the past few years, several similarities have been observed in the X-ray variability of nearby AGN and XRBs, supporting the idea of {\it black hole unification} (M$^{c}$Hardy et al. 2006). An important part of this project is the reliable estimation of  power spectra from AGN, which are expected to vary like X-ray binaries but slower by a factor of $\sim 10^{5-8}$, commensurate with their much larger black hole masses  (e.g. Shakura \& Sunyaev 1976; Mushotzky, Done \& Pounds 1993). The timescales of interest for AGN variability studies therefore range from $\sim 100$ sec to $\gsim$years, which leads to data analysis problems that are usually not present for the fast X-ray binary data. 

If we take the power spectrum of Cygnus X-1 (Figure \ref{fig:cyg}) and scale the frequencies by a factor $\lsim 10^{-5}$ to those expected for an AGN, we would expect to see the power spectrum as `red' (i.e. increasing to lower frequencies like $f^{-\alpha}$ with $\alpha \gsim 1$) down to frequencies as low as $\sim 10^{-6}$ Hz (timescales of weeks), and there are good reasons for expecting the red spectrum to extend much lower. The sources also tend to have much lower count rates (fluxes) and so the Poissonian fluctuations in the count rate (measurement error) is relatively stronger. Clearly an observation lasting only $\sim$hour, like that shown in Figure \ref{fig:dps}, will not be sufficient to obtain a good power spectrum estimate for a much fainter, slower AGN. On the timescales of most observations we are looking at `long memory' processes when we observe AGN, with significant power outside the observed frequency bandpass. 

\begin{figure}
  \centering
	\includegraphics[width=13.0cm, angle=0]{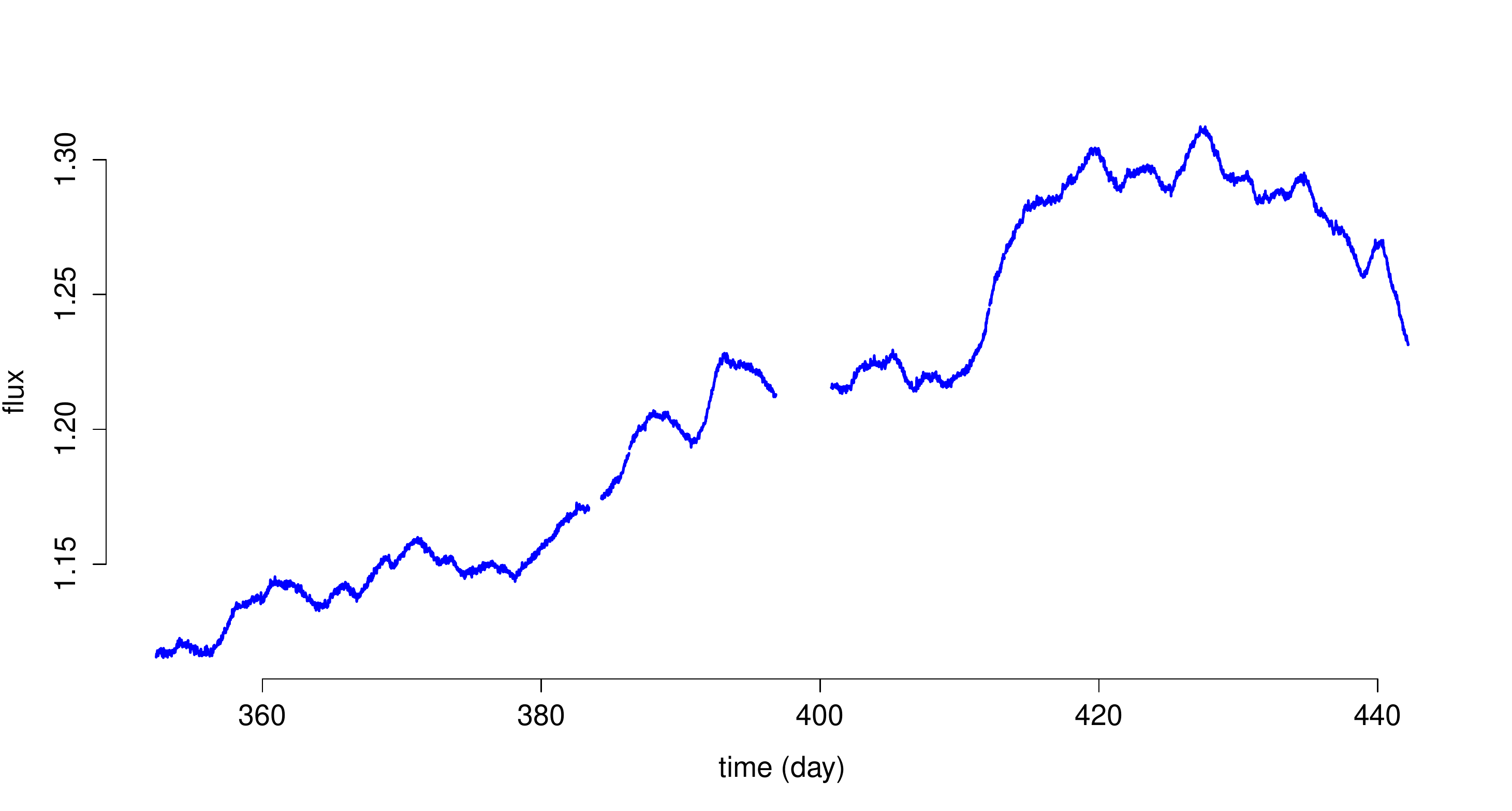} 
\caption{A section of the {\it Kepler} light curve of the AGN Zw $229-15$ from Mushotzky et al. (2011). The series contains $N=4125$ points over $90$ days, sampled every $29.4$ minutes with the exception of a few short gaps, and the measurement errors on the flux are $\lsim 0.1$\%. Clearly the variations are dominated by longest timescales or lowest frequencies, i.e. this is a realisation of a red noise (long memory) process with a steep power spectrum.}
\label{fig:kepler}
\end{figure}

In order to obtain a useful power spectrum estimate we need low variance, from large $M$, equivalent to many repeated samples at each frequency, and low bias, from long segments, i.e. large $N$, and good time sampling $\Delta t$, covering the full range of frequencies that show an interesting (non-white) spectrum.
But it is almost never practical to observe an AGN continuously for many, many years, at high time resolution (e.g. seconds-minutes), to meet these criteria. AGN time series come from `monitoring campaigns' comprising regular short snapshot observations of a target extending over a long period, or continuous but much shorter `long look' observations, or combinations of the two. The monitoring approach often suffers from the effects of {\it aliasing} of high frequency power (Uttley, M$^{c}$Hardy \& Papadakis 2002; Kirchner 2005), while the shorter, intensive observations often suffer from {\it leakage} of lower frequency power (Deeter \& Boynton 1982; Uttley et al. 2002). In both cases significant power from frequencies outside the observed frequency bandpass biases the results of simple spectral estimators, which can be understood in terms of the effect of the window function (the time sampling pattern) on the Fourier transform of the time series. Figure \ref{fig:kepler} shows a rare exception to the rule that long timescale monitoring data are sparsely sampled: it shows data from the {\it Kepler} mission that continuously monitors the brightness of thousands of sources in a single patch of sky. Even with such good data there is a severe problem of spectral leakage of power from low frequencies.

It is a fundamental fact of spectral analysis that the window function biases power spectrum estimates (e.g. Priestley 1981; Chatfield 2004), but this bias should disappear in the limit of large $N$ (and small $\Delta t$) for stationary processes. The expectation value of the periodogram $I_j$ at Fourier frequency $f_j$ is 
\begin{equation}
  E[I_j] = \int_{-f_{Nyq}}^{+f_{Nyq}} F(f_j-f^{\prime}) S(f^{\prime}) df^{\prime} = S(f) - b_I(f)
\end{equation}
where $S(f)$ is the true spectral density function, $F(f)$ is the Fejer kernel, and $f_{Nyq} = 1/2\Delta t$ is the Nyquist frequency. The above is valid when there is no significant power in the spectrum above $f_{Nyq}$ -- i.e. the variations are well resolved by sampling at discrete times $\Delta t$. If there is significant high frequency power above $f_{Nyq}$ then additional terms (power from high frequencies, called {\it aliases}) also bias the periodogram.

The convolution by the Fejer kernel distorts the periodogram away from the true spectrum, and results from the finite duration of the time series (see Priestley 1981, chapter 6). The expected difference between the true spectrum and the periodogram is the bias on the periodogram: $b_I(f_j) = E[S(f_j) - I_j]$. 
The Fejer kernel depends on the time series duration, $T = N \Delta t$, and has a main lobe of width $\Delta f \sim 1/T$ plus oscillatory side-lobes that decay as $\sim 1/f^2$ either side. It is these side-lobes that cause the leakage -- the transfer of power between distant Fourier frequencies. 

The basic problem in AGN timing analysis is that we cannot obtain sufficiently long and well-sampled time series to give negligible bias. This also means there are not enough data to average over $M$ non-overlapping intervals to give Normally distributed averages. The time series literature contains many discussions of power spectral estimation for `long memory' processes (e.g. Fougere 1985; Hurvich \& Beltrao 1993; Beran 1993; McCoy, Walden \& Percival 1998). Popular methods often involve some kind of {\it tapering} of the time series, a modification of the window function to reduce leakage, as in the Multiple Taper Method (MTM; McCoy et al. 1998 and references therein), or {\it differencing}, i.e. examination of $\Delta(t) = x(t) - x(t-1)$, rather than $x(t)$. 

\subsection{The forward problem}

The approach being developed by astronomers to solve this problem is based on `forward fitting.' Rather than trying to deconvolve the effects of the window function from the power spectrum estimate (the {\it inverse problem}), we include the window function biases in model power spectra that can be compared to the data (without tapering etc.) The approach could be said to be {\it parametric}, being based on comparison of the data with specific well-defined models, rather than the {\it non-parametric} approach of deconvolution, but this is reasonable as the goal of much of this work is to compare physical models to data. The `response' method (Uttley,  M$^{c}$Hardy \& Papadakis 2002 based on earlier work of Done et al. 1992 and Green, M$^{c}$Hardy \& Lehto 1999) relies heavily on simulations. The essence of the method is as follows:
\begin{itemize}
\item
Estimate the power spectrum of the data $x(t_i)$, $i=1,2,\ldots,N$ using e.g. the periodogram (with optional smoothing/averaging)
\item
define a power spectrum model $S(f; \theta)$ in terms of parameters $\theta$
\item Repeat $N_{\rm sim}$ times:
\begin{itemize}
\item
generate a random time series based on $S(f; \theta)$
\item
sample the simulated series in the same manner as the real data (i.e. extract a subset from times $t_i$)
\item
compute the power spectral estimate of the sampled simulation
\end{itemize}
\item
at each frequency compute the mean and variance-covariance of the $N_{\rm sim}$ estimates
\item
compare the mean spectral estimates from simulations to those from the real data
\end{itemize}
The simulation step needs to be handled carefully. Usually the algorithm of Timmer \& K\"{o}nig (1995)\footnote{
This algorithm seems to have been discovered several times. Very similar methods are discussed by Ripley (1987; section 4.5), Davies \& Harte (1987), and Davis et al. (1981).} is used to form random data in Fourier space that are then inverse Fourier transformed into a random time series. It is usually necessary to produce simulations much longer, and with finer time sampling, than the real data, from which a subset of the points are extracted in order to simulate the process of sampling a limited, discrete time series from a continuous process. Furthermore, simulations produced in this manner are realisations of stationary, linear, Gaussian processes. Non-linearities such as the `rms-flux' effect discussed above can be included in the simulation if necessary (Uttley et al. 2005).

There is still some discussion about how best to perform the statistical comparison of data and simulations -- e.g. how to define a fit statistic to use when searching for the best fitting model and its parameters. (It is not trivial to apply e.g. the {\it Whittle} likelihood in the presence of leakage and/or aliasing biases.) But the main advantage of this approach is that it should account correctly for the effect of the finite and discrete time sampling on the distribution of the power spectral estimates at each frequency (aliasing and leakage), by `folding' the model spectra though the same sampling process. Other observational effects such as measurement errors or non-linear detector response can also be included in the simulation procedure. The main disadvantage is computing time -- a large number of (potentially long) time series simulations needs to be performed for each choice of model and parameter values $\theta$.

This approach has proved quite powerful for extracting reliable information about power spectral models from time series of (relatively) slowly varying AGN that are (relatively) poorly sampled. But work remains to be done. Can this simulation-based approach be replaced by a faster analytical equivalent? What statistic(s) should be used for data-model comparison? Can non-stationarity be taken into account? There is also a problem with long-term monitoring campaigns in that the data are usually not very evenly sampled in time, yet standard Fourier methods (and many time domain methods) are best understood for evenly sampled time series. Generalisations of some methods to unevenly sampled data, such as the Lomb-Scargle periodogram (Scargle 1982), have been developed for special cases, such as searching for sinusoidal signals in white noise. To my knowledge the effect of uneven sampling on the recovery of red power spectra remains an open problem (e.g. Milotti 2007; Stahn \& Gizon 2008).


\section{Multivariate time series}
\label{sect:bi}

Bivariate, or more generally multivariate time series offer a new dimension to the analysis and potential for model testing. In many cases the only information we can obtain on astrophysical phenomena is contained in the time and wavelength distribution of the light we receive. We might have observations from two telescopes operating simultaneously in different wavebands (e.g. optical and X-rays), or from one detector that tags events according to their energy (wavelength) so that we can form time series in different ranges of photon energy (wavelength). Depending on the situation the analysis of such data falls under different names: often this might be called {\it spectral variability} meaning the time variability of the photon energy/wavelength spectrum (not the power spectrum!). Another situation is when we receive light from multiple images of the same event and form a time series from each image, such as with gravitationally lensed quasars (Courbin 2003).

In some cases having data in both time and wavelength dimensions allow for the recovery of spatial information from objects so distant they appear only as points in our best images. For example, the periodic rotation of single stars or the revolution of binary stars about their centre of mass can produce periodic variations in the light received on Earth when there is anisotropic emission such as star spots or a structured accretion disc.  The rotation of these systems causes periodic modulations in wavelengths of atomic features (the relativistic Doppler shift), and by resolving the wavelength (and hence radial velocity) changes as a function of the phase of the period it is possible to extract information about the spatial structure by the method of Doppler tomography (Marsh 2001). This can be viewed as a problem in multivariate time series analysis where we observe simultaneous time series for each of many distinct Doppler velocities.

\begin{figure}
  \centering
	\includegraphics[width=13.0cm, angle=0]{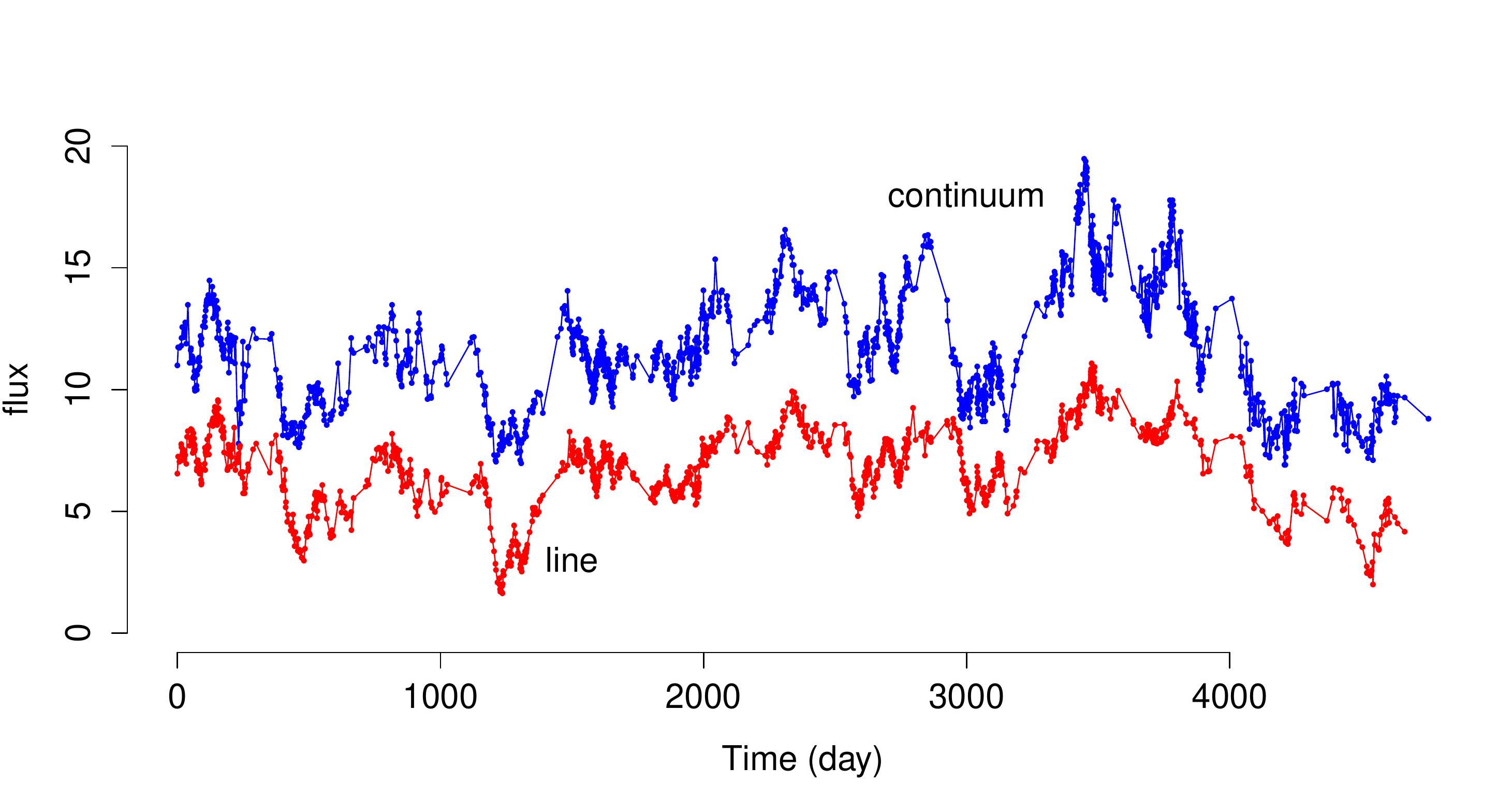} 
\caption{Data from a `reverberation mapping' campaign organised by the `AGN Watch' consortium. The two time series are based on repeated observations of the active galaxy NGC 5548 taken over 1988--2001 (Peterson et al. 2002), and represent part of the optical continuum flux (around $5100$ \AA) and a line flux (the Hydrogen Balmer H$\beta$ line), where the line is expected to respond to the continuum with some time delay (an `echo') resulting from the size of the line emitting region. Notice the irregular sampling that is inevitable for such long timescale monitoring from the ground. In this case there is a delay of $\sim 20$ days between the driving continuum variations and the line response, but this delay is not constant through the $13$ years of data.}
\label{fig:agn}
\end{figure}

A related method is that of `reverberation mapping' of AGN (Peterson 1993). AGN often show strong atomic emissions lines, such as the Hydrogen Balmer series lines, excited by photoionisation and recombination from the centrally generated continuum emission. The lines are therefore expected to respond to variations in the luminosity of the ionising continuum, and time delays between continuum and line variations relate to the light travel time (and hence distance) between continuum and line emitting regions. More specifically, the goal is to recover the {\it impulse response}\footnote{In some astronomy contexts this is also called the {\it transfer function}.} of the line (its response as a function of time to an instantaneous impulse) and relate this to the spatial distribution of the emitting gas. If $x(t)$ is the driving continuum, then the line emission $y(t)$ is given by
\begin{equation}
\label{eqn:transfer}
y(t)=\int_{-\infty}^{+\infty} \psi(t - \tau) x(\tau) d\tau = \psi(t) \otimes x(t).
\end{equation}
where $\psi(t)$ is the  response function. As with Doppler tomography, the time variations are used to probe unresolved spatial structure. But AGN do not vary periodically like rotating or binary stars and so the delays are seen as `echoes' of the variations of a (red) noise process. Figure \ref{fig:agn} shows an example of data obtained from a sustained reverberation mapping campaign. The recovery of the time response of the emission line is hampered by the irregular time sampling and the very red (low frequency dominated) nature of the continuum variations, and the fact that the line response is likely to be non-linear and non-stationary at some level.

The time delays between correlated time series of different but related origin tell us about the causal connections within the system. In reverberation mapping observations we see the continuum increase in flux, and the line respond in a correlated manner with some (reproducible) delay. In some cases we do not understand the physical relationship between different parts of the wavelength spectrum of sources and so correlation, time delays, and multivariate time series methods in general, are used to uncover these causal links.

\subsection{Cross-correlation and cross-spectrum methods}

As with most time series problems one can consider the problem in the time domain or the frequency (Fourier) domain. The primary analysis tool for comparing two time series in the time domain is the cross-correlation or cross-covariance function (Priestley 1981; Chatfield 2004). If we have two series $x_t$ and $y_t$, the cross-covariance is defined as 
\begin{equation}
 \gamma_{xy}(k) = E[ (x_t - E[x_t]) (y_{t-k} - E[y_t]) ],
\end{equation}
and the cross-correlation is $\rho_{xy}(k) = \gamma_{xy}(k) / \{ \gamma_{xx}(0) \gamma_{yy}(0) \}^{1/2} = \gamma_{xy}(k) / \sigma_x \sigma_y$. The {\it sample} cross-covariance is usually simple to calculate for long, evenly sampled, simultaneous time series. ($E$ is the expectation.)

But when the two time series are not evenly sampled, perhaps not even sampled simultaneously, and are contaminated by measurement errors, it is less obvious how to perform the calculation. The solutions used by astronomers typically involve some kind of interpolation of one or both time series to evenly spaced and/or simultaneous times, or averaging together pairs of point with similar time differences (see e.g. White \& Peterson 1994). 
Even when the issue of computation is resolved there remain questions for interpreting the cross-correlation: when is a peak an indication of a meaningful correlation between $x_t$ and $y_t$ if these are non-white noise processes? how should we estimate the time delay and its confidence intervals? or more generally, how do we recover the  response function? Here, as with power spectrum estimation, simulation-based Monte Carlo methods are proving useful (e.g. Uttley et al. 2003).

The cross-spectrum is a related tool used for comparing the properties of two simultaneous time series in the frequency domain (Prisetley 1981; Chatfield 2004).
It is the Fourier transform of the cross-covariance\footnote{The cross-spectrum and cross-covariance are Fourier counterparts, just as the power spectrum and auto-covariance are Fourier counterparts.}, or the product of the Fourier transforms of the two light curves: $C(f) = X^{\ast}(f) Y(f)$.  In principle at least, the cross-spectrum contains the same information as the cross-covariance. But the cross-spectrum can often be computed quickly (using a Fast Fourier Transform), and can be a more direct route to the  response function relating the two time series. The cross spectrum is complex valued, represented in the complex plane by an amplitude and a phase. If we write the complex Fourier transforms of the two light curves as $X(f) = |X(f)|{\rm e}^{{\rm i} \phi_x(f)}$ and $Y(f) = |Y(f)| {\rm e}^{{\rm i} \phi_y(f)}$ then the cross spectrum can be written as:
\begin{equation}
\label{eqn:cross_spec2}
C(f) = |X(f)| |Y(f)| {\rm e}^{{\rm i} (\phi_y(f) - \phi_x(f))} = |C(f)| {\rm e}^{{\rm i} \Delta \phi(f)}
\end{equation}
In practice, the estimation of the cross-spectrum from a finite sample of real data usually requires a more involved procedure, see e.g. Vaughan \& Nowak (1997). Usually the cross-spectrum is divided into its modulus and phase. The phase component, $\Delta \phi(f) = \arg[C(f)]$ represents the phase difference between the two light curves (which corresponds to a time delay $\tau(f) = \Delta \phi(f) / 2 \pi f$ at that frequency). The squared magnitude of the cross spectrum $|C(f)|^2$ is used to define the {\it coherence} of the two light curves (different from the {\it coherence} of a peak in a power spectrum). The cross-spectral phase and coherence have been used to great effect on the evenly sampled X-ray data from sources like Cygnus X-1 (Miyamoto \& Kitamoto 1989; Nowak et al. 1999), revealing complex, frequency-dependent time delays between X-rays of different energies. Recently such methods have provided evidence for the causal connections between different parts of the accretion flow in X-ray binaries and in AGN (Uttley et al. 2011; Fabian et al. 2009). 

In order to illustrate the cross-spectrum, consider two time series $x(t)$ and $y(t)$, evenly and simultaneously sampled, that are related by a  response function (equation \ref{eqn:transfer}). From the convolution theorem we have
\begin{equation}
\label{eqn:transfer2}
Y(f) = \Psi(f) X(f),
\end{equation}
where $\Psi(f)$ is the Fourier transform of $\psi(t)$. Therefore:
\begin{equation}
C_{xy}(f)= X^{\ast}(f) Y(f)  = X^{\ast}(f) \Psi(f) X(f) = \Psi(f) |X(f)|^2 = \Psi(f) S_x(f)
\end{equation}
where $S_x(f) = |X_f|^2$ is the power spectrum of $x(t)$ (neglecting a normalising factor). Estimates of the cross-spectrum $C_{xy}$ and power spectrum $S_x$ therefore lead to a simple estimate of the response $\Psi(f)$. If the two light curves are related by a single, linear response function (e.g. a delay or a smoothing) then they will have unity coherence (in expectation and in the absence of bias). Non-linear relations and partially independent variations result in lower coherence.

Given $L$ evenly sampled time series (corresponding to channels $l=1,2,\ldots,L$) one could compute $L(L-1)/2$ cross-spectra, but in most cases there will be a great deal of redundancy between these. The phase delay (at a given frequency) between channels $1$ and $3$ cannot be independent of the phase delays between channels $1$ and $2$ and between channels $2$ and $3$. There will usually be fewer response functions relating the $L$ bands than pairs of time series to form cross-spectra (Vaughan 1997). Another approach being explored by e.g. Uttley et al. (2011) is to form only $L$ cross-spectra by comparing the time series for each channel $l$ against the summed time series of all the other channels. This has some utility when the channels are adjacent in wavelength and the number of underlying varying components to the emission is small compared to $L$. 

Currently the frequency domain approach (cross-spectrum) is applied to data that are evenly sampled, like those from short, intensive X-ray observations, whereas the time domain approach (cross-covariance) is often applied to unevenly sampled data like those from long-term AGN monitoring. But these tools may not be optimal for those tasks. Some outstanding problems in this area are: how best to compute reliable time delays (and, more generally, response functions) from unevenly and non-simultaneously sampled light curves? how best to recover transfer functions from very red time series (where again spectral leakage can bias the results)? how best to model multi-channel time series in terms of their simultaneous power spectra and cross-spectra (or time domain equivalents)?


\subsection{Time domain methods}

It seems to be the case that a great deal of astronomical time series analysis is based around spectral methods, rather than time domain methods. This choice largely motived by scientific and practical considerations. There may also be historical and sociological factors involved. But I certainly do not wish to give the impression that only spectral methods are used in astronomy. For example, Scargle (1981, 1998) discussed in great detail the time domain modelling of random time series in astronomy. Kelly et al. (2010) and Miller et al. (2010) developed tools for recovering the power spectra of AGN by fitting to the data in the time domain.
Press \& Rybicki (1998) and Zu et al. (2011) developed time domain methods for reconstructing missing data from irregularly sampled AGN time series in order to 
recover time delays from pairs of unevenly sampled time series (the method is similar to the Kalman filter). Another method often used on unevenly sampled data is the {\it structure function}, although Emmanoulopoulos et al. (2010) give strong arguments against its value as an analysis tool. 


\section{Concluding remarks}
\label{sect:conc}

All the data presented in this paper have been obtained from public archives. The {\it RXTE} archive contains many thousands of observations of rapidly varying X-ray binaries and slowly varying AGN; the {\it Swift} archive contains data on gamma-ray, X-ray and ultraviolet variability of gamma-ray bursts (GRBs) but also a host of other transient and persistent sources; and time series data from other wavebands for thousands of sources are being made available right now (e.g. the public {\it Kepler} archive). I encourage readers to download some data and explore the huge range of data and physics available to astronomical time series analysis. It may be that we are just beginning to explore the science possible with astronomical time series data (Lawrence 2007).


\section*{Acknowledgements}

I would like to thank the meeting organisers (Nick Jones and Tom Maccarone) for allowing him to submit the manuscript rather late. 
I also extend thinks to Phil Uttley and the two referees for their valuable comments on earlier versions of this manuscript.
This research has made use of NASA's Astrophysics Data System.



\begin{thebibliography}

\bibitem{bib1} Appourchaux, T., 2011 A crash course on data analysis in asteroseismology. (arXiv:1103.5352)

\bibitem{bib2} Arnaud, K., Smith, R., Siemiginowska, A,. 2011, Handbook of X-ray Astronomy. Cambridge: Cambridge University Press

\bibitem{bib3} Auvergne M., et al. 2009 The CoRoT satellite in flight: description and performance. \textit{Astro. Astroph} \textbf{506} 411--424

\bibitem{bib4} Barret, D., et al. 2005 On the high coherence of kHz quasi-periodic oscillations. \textit{Mon. Not. Roy. Ast. Soc.} \textbf{357} 1288--1294

\bibitem{bib5} Barret, D., Vaughan S. 2012, Maximum Likelihood Fitting of X-Ray Power Density Spectra: Application to High-frequency Quasi-periodic Oscillations from the Neutron Star X-Ray Binary 4U1608-522. \textit{Astroph J.} \textbf{746}:131

\bibitem{bib7} Bartlett, M. S. 1948 Smoothing Periodograms from Time-Series with Continuous Spectra. \textit{Nature} \textbf{161}, 686--687

\bibitem{bib8} Belloni, T. 2007, Noise components from black-hole binaries in our galaxy. \textit{Proc. SPIE} \textbf{6603} 660311 (arXiv:0802.0376)

\bibitem{bib11} Beran, J., 1993, Fitting long-memory models by generalized linear regression. \textit{Biometrica} \textbf{80} 817--822

\bibitem{bib13} Bernardini, E. 2011 Astronomy in the Time Domain. \textit{Science} \textbf{331}, 686--687

\bibitem{bib15} Borucki, W. J., et al. 2010 Kepler Planet-Detection Mission: Introduction and First Results. \textit{Science} \textbf{327} 977--980

\bibitem{bib17} Brillinger,  D. R., Rosenblatt, M., 1967 Computation and interpretation of the $k$-th order spectra. In \textit{Spectral
Analysis of Time Series} (eds. B. Harris) pp. 189--232. New York: Wiley

\bibitem{bib20} Burrows, D. N. et al. 2011 Relativistic jet activity from the tidal disruption of a star by a massive black hole \textit{Nature} \textbf{476}, 421-–424

\bibitem{bib201} Chatfield, C., 2004 The Analysis of Time Series: An Introduction (6th ed). London: Chapman \& Hall/CRC

\bibitem{bib21} Courbin, F., 2003 Quasar Lensing: the Observer's Point of View (arXiv:astro-ph/0304497)

\bibitem{bib23} Davies, R. B., Harte, D. S., 1987 Tests for Hurst effect. \textit{Biometrika} \textbf{74} 95--101

\bibitem{bib24} Davis, B. M., Hagan, R. and Borgman, L. E., 1981, A Program for the Finite Fourier Transform Simulation of Realizations from a One-Dimensional Random Function with Known Covariance. \textit{Comp. and Geosci.}, \textbf{7}, 199–206

\bibitem{bib25} Deeter, J. E., Boynton, P. E., 1982 Techniques for the estimation of red power spectra. I - Context and methodology. \textit{Astroph. J.} \textbf{261} 337--350

\bibitem{bib30} Done, C., et al. 1992 The X-ray variability of NGC 6814 - Power spectrum. \textit{Astroph. J.} \textbf{400}  138--152

\bibitem{bib33} Emmanoulopoulos, D., M$^{c}$Hardy, I. M., Uttley, P., 2010, On the use of structure functions to study blazar variability: caveats and problems. \textit{Mon. Not. Roy. Ast. Soc.} \textbf{404} 931--946

\bibitem{bib36} Fabian, A. C., et al., 2009 Broad line emission from iron K- and L-shell transitions in the active galaxy 1H0707-495. \textit{Nature} \textbf{459} 540--542

\bibitem{bib38} Fackrell, J., McLaughlin S., 1996 Detecting nonlinearities in speech sounds using the bicoherence. \textit{Proceedings of the Institute of Acoustics} \textbf{18(9)} 123--130

\bibitem{bib46} Fender, R., Belloni, T. 2004 GRS 1915+105 and the Disc-Jet Coupling in Accreting Black Hole Systems. \textit{Ann. Rev. Astr. Astroph.} \textbf{42} 317--364



\bibitem{bib50} Fender, R. 2011 The scientific potential of LOFAR for time-domain astronomy. \textit{ArXiv e-prints} arXiv:1112.2580

\bibitem{bib60} Feroci, M., 2011 The Large Observatory for X-ray Timing (LOFT). \textit{Exp. Astron.} DOI: 10.1007/s10686-011-9237-2 (arXiv:1107.0436)

\bibitem{bib65} Feigelson, E. D., 1997 Time series problems in astronomy: an introduction. In \textit{Applications of Time Series Analysis in Astronomy and Meteorology} (eds. T. Subba Rao, M.B. Priestley, O. Lessi). pp. 161--186. London: Chapman \& Hall

\bibitem{bib70} Ford E. C., 2011 Transit Timing Observations from Kepler. I. Statistical Analysis of the First Four Months. \textit{Astroph J. Suppl.} \textbf{197} 2

\bibitem{bib73} Fougere, P. F., 1985 On the accuracy of spectrum analysis of red noise processes using maximum entropy and periodogram methods: simulation studies and application to geophysical data. \textit{J. Geophys. Res.} \textbf{90} 4355--4366

\bibitem{bib75} Gehrels, N., Ramirez-Ruiz, E., Fox D. B. 2009 Gamma-Ray Bursts in the Swift Era. \textit{Ann. Rev. Astr. Astroph.} \textbf{47}, 567-–617

\bibitem{bib78} Gregory, P. C., 2011 Bayesian exoplanet tests of a new method for MCMC sampling in highly correlated model parameter spaces. \textit{Mon. Not. Roy. Ast. Soc.} \textbf{410} 94--110 

\bibitem{bib779} Green, A. R., M$^{c}$Hardy, I. M., Done, C., 1999 The discovery of non-linear X-ray variability in NGC 4051. textit{Mon. Not. Roy. Ast. Soc.} \textbf{305}  309--318

\bibitem{bib79} Heil, L. M., Vaughan, S., Uttley, P. 2012 The Ubiquity of the rms-flux relation in Black Hole X-ray Binaries. \textit{Mon. Not. Roy. Ast. Soc.} \textbf{in press} (arXiv:1202.5877)

\bibitem{bib790} Hurvich, V. M., Beltrao, K. I., 1993 Asymptotics for the low-frequency ordinates of the periodogram of a long-memory time series. \textit{J. Time Series Analysis} \textbf{14}  455--472

\bibitem{bib798} Izenman, A. J., 1983 An historical note on the Zurich sunspot relative numbers. \textit{J. Roy. Stat. Soc. A} \textbf{146} 311--318

\bibitem{bib80} Kaiser, N. et al. 2002 Pan-STARRS: A Large Synoptic Survey Telescope Array. \textit{Proc. SPIE} \textbf{4836} 154

\bibitem{bin85} Kelly, B. C., Sobolewska, M., Siemiginowska, A., 2011, A Stochastic Model for the Luminosity Fluctuations of Accreting Black Holes. \textit{Astroph J.}
 \textbf{730} 52

\bibitem{bib90} Kim, Y. C., Powers, E. J., 1979 Digital bispectral analysis and its applications to nonlinear wave interactions.
\textit{IEEE Transactions on Plasma Science} \textbf{7} 120--131

\bibitem{bib95} Kirchner, J. W., 2005 Aliasing in 1/f noise spectra: Origins, consequences, and remedies. \textit{Phys. Rev. E}, \textbf{71}, 066110

\bibitem{bib100} Kulkarni, S. R. 2011 Cosmic Explosions (Optical Transients), \textit{ArXiv e-prints} arXiv:1202.2381

\bibitem{bib105} Lasota, J.-P., 2001, The disc instability model of dwarf novae and low-mass X-ray binary transients. \textit{New Astr. Rev.} \textbf{45} 449--508

\bibitem{bib110} Lawrence, A., 2007 Wide-field surveys and astronomical discovery space. \textit{Astron. \& Geophys.} \textbf{48} 3.27--3.33

\bibitem{130} Liu, K., et al. 2011 Prospects for high-precision pulsar timing. \textit{Mon. Not. Roy. Ast. Soc.} \textbf{417} 2916--2926

\bibitem{bib160} Maccarone T. J., Coppi, P. S., 2002 Higher order variability properties of accreting black holes. \textit{Mon. Not. Roy. Ast. Soc.} \textbf{336} 817--825

\bibitem{bib165} Maccarone T. J. et al. 2011 Coupling between QPOs and broad-band noise components in GRS $1915+105$. \textit{Mon. Not. Roy. Ast. Soc.} \textbf{413} 1819-1827

\bibitem{bib178} Marsh, T. R. 2001 Doppler Tomography. In \textit{Astrotomography, Indirect Imaging Methods in Observational Astronomy}, eds H. M. J. Boffin, D. Steeghs, J. Cuypers, Lecture Notes in Physics \textbf{573} 1--26

\bibitem{bib180} Marsh, T. R. 2008 High-Speed Optical Spectroscopy. In {\it High Time Resolution Astrophysics} (eds.  D. Phelan, O. Ryan, and A. Shearer) pp. 75--93. Berlin: Springer

\bibitem{bib190} McCoy, E.J., Walden, A.T., Percival, D.B., 1998 Multitaper spectral estimation of power law processes. \textit{IEEE Trans. Signal Processing} \textbf{46} 655--668

\bibitem{bib200} Mendez, M. et al., 1998 Discovery of a Second KHZ QPO Peak in 4U 1608-52. \textit{Astroph J.} \textbf{494}, L65--L69

\bibitem{bib205} Miller, L., et al., 2010, Spectral variability and reverberation time delays in the Suzaku X-ray spectrum of NGC 4051. \textit{Mon. Not. Roy. Ast. Soc.} \textbf{403}, 196--210

\bibitem{bib210} Milotti, E., 2007 Artifacts with uneven sampling of red noise. \textit{Phys. Rev. E} \textbf{75} 011120

\bibitem{bib215} {Miyamoto}, S. and {Kitamoto}, S., 1989 X-ray time variations from Cygnus X-1 and implications for the accretion process. \textit{Nature} \textbf{342} 773--774

\bibitem{bib220} {Mushotzky}, R.~F.,  {Done}, C., {Pounds}, K.~A.,  1993 X-ray spectra and time variability of active galactic nuclei. \textit{Ann. Rev. Astr. Astroph.} \textbf{31}, 717--761

\bibitem{bib221} {Mushotzky}, R.~F., et al. 2011 Kepler Observations of Rapid Optical Variability in Active Galactic Nuclei. \textit{Astroph. J.} \textbf{743} L12

\bibitem{bib250} Nikias, C. L., Petropulu, A. P., 1993 {\it Higher-order spectra analysis - A nonlinear signal processing framework}.
New Jersey: Prentice-Hall

\bibitem{bib260} Nowak, M. A. et al. 1999 Rossi X-Ray Timing Explorer Observation of Cygnus X-1. II. Timing Analysis. \textit{Astroph. J.} \textbf{520}  874--891

\bibitem{bib268} Peterson, B. M. 1993 Reverberation mapping of active galactic nuclei. \textit{Pub. Astron. Soc. Pacific} \textbf{105} 247--268

\bibitem{bib270} Peterson, B. M. et al. 2002 Steps toward Determination of the Size and Structure of the Broad-Line Region in Active Galactic Nuclei. XVI. A 13 Year Study of Spectral Variability in NGC 5548. \textit{Astroph. J.} \textbf{581} 197--204

\bibitem{bib276} Priestley, M. B., 1981 Spectral analysis and time series. London: Academic Press

\bibitem{bib280} Press W. H., Rybicki, G. B., 1998, Magnification Ratio of the Fluctuating Light in Gravitational Lens 0957+561. \textit{Astroph. J.} \textbf{507} 108-112

\bibitem{bib290} Rial, J., Anaclerio, C. 2000 Understanding nonlinear responses of the climate system to orbital forcing.
\textit{Quaternary Science Reviews} \textbf{19} 1709--1722

\bibitem{bib293} Ripley, B. D., 1987 Stochastic Simulation, New York: Wiley

\bibitem{bib294} Scargle, J. D., 1981 Studies in astronomical time series analysis. I - Modeling random processes in the time domain. \textit{Astroph. J. Supp.} \textbf{45} 1--71

\bibitem{bib295} Scargle, J. D., 1998 Studies in astronomical time series analysis. IV - Bayesian Blocks, a New Method to Analyze Structure in Photon Counting Data \textit{Astroph. J.} \textbf{504} 405--418

\bibitem{bib298} Scaringi, S., et al., 2012, The universal nature of accretion-induced variability: the rms-flux relation in an accreting white dwarf. \textit{Mon. Not. Roy. Ast. Soc.} \textbf{421} 2854--2860

\bibitem{bib300} {Shakura}, N.~I.,  {Sunyaev}, R.~A.,  1976 \textit{Mon. Not. Roy. Ast. Soc.} \textbf{175} 613--632

\bibitem{bib310} Stahn, T., Gizon, L., 2008 Fourier analysis of gapped time series: Improved estimates of solar and stellar oscillation parameters. \textit{ Solar Phys.} \textbf{251} 31--52

\bibitem{bib320} Stella, L., Vietri M. 1999 kHz Quasiperiodic Oscillations in Low-Mass X-Ray Binaries as Probes of General Relativity in the Strong-Field Regime. \textit{Phys. Rev. Lett.} \textbf{82} 17--20

\bibitem{bib340} Strohmayer, T., Bildsten, L., 2006, New views of thermonuclear bursts. In {\it Compact stellar X-ray sources} (eds. W. Lewin and M. van der Klis) pp113--156. Cambridge: Cambridge University Press

\bibitem{bib350} Timmer, J., K\"{o}nig, M., 1995, On generating power law noise. \textit{Astron \& Astroph.} \textbf{300} 707--710

\bibitem{bib360} Uttley, P. M$^{c}$Hardy, I. M., Papadakis, I. 2002  Measuring the broad-band power spectra of active galactic nuclei with RXTE \textit{Mon. Not. Roy. Ast. Soc.} \textbf{332} 231--250

\bibitem{365} Uttley, P. Edelson, R., M$^{c}$Hardy, I. M., Peterson, B. M., Markowitz, A., 2003, Correlated Long-Term Optical and X-Ray Variations in NGC 5548. \textit{Astroph. J.} \textbf{584} L53--L56

\bibitem{bib370} Uttley, P. M$^{c}$Hardy, I. M., Vaughan, S. 2005 Non-linear X-ray variability in X-ray binaries and active
galaxies. \textit{Mon. Not. Roy. Ast. Soc.} \textbf{359} 345--362

\bibitem{bib375} Uttley, P., et al. 2011 The causal connection between disc and power-law variability in hard state black hole X-ray binaries. \textit{Mon. Not. Roy. Ast. Soc.} \textbf{414} L60--L64

\bibitem{bib400} van der Klis, M. 1989 Fourier techniques in X-ray timing. In {\it Timing Neutron Stars} (eds.  H. \"{O}gelman and E. P. J. van den Heuvel) pp. 27--69.  New York: Kluwer Academic / Plenum Publishers

\bibitem{bib405} van der Klis, M. 2006 Rapid X-ray Variability. In {\it Compact stellar X-ray sources} (eds. W. Lewin, M. van der Klis) pp. 39--112. Cambridge, UK: Cambridge University Press

\bibitem{bib406} van der Klis, M. 2007 Millisecond phenomena in mass accreting neutron stars. \textit{Proc. SPIE} \textbf{6603} 660312 

\bibitem{bib440} Vaughan, B., 1997 A multi-channel, cross-spectral technique for calculating best-fit time-delay spectra. In \textit{Applications of Time Series Analysis in Astronomy and Meteorology} (eds. T. Subba Rao, M.B. Priestley, O. Lessi). pp. 264--274. London: Chapman \& Hall

\bibitem{bib441} Vaughan, B., Nowak, M. A., 1997 X-Ray Variability Coherence: How to Compute It, What It Means, and How It Constrains Models of GX 339-4 and Cygnus X-1. \textit{Astroph. J.} \textbf{474} L43--L46

\bibitem{bib450} Vaughan, S., et al. 2003 On characterizing the variability properties of X-ray light curves from active galaxies. \textit{Mon. Not. Roy. Ast. Soc.} \textbf{345} 1271--1284.

\bibitem{bib500} Vaughan, S., Uttley, P. 2007, Studying accreting black holes and neutron stars with time series: beyond the power spectrum. \textit{Proc. SPIE} \textbf{6603} 660314 (arXiv:0802.0391)

\bibitem{bib600} White, R. J., Peterson, B. M., 1994 Comments on cross-correlation methodology in variability studies of active galactic nuclei. \textit{Pub. Astron. Soc. Pacific} \textbf{106}  879--889

\bibitem{bib700} Zu, Y. Kochanek, C. S., Peterson, B. M., 2011, An Alternative Approach to Measuring Reverberation Lags in Active Galactic Nuclei. \textit{Astroph. J.} \textbf{735} 80

\end{thebibliography}
\end{document}